\documentclass[twocolumn]{aastex63}

\usepackage{graphicx}
\usepackage[T1]{fontenc}
\usepackage{aecompl}
\usepackage[]{amsmath}

\usepackage{color}

\usepackage{xspace}

\usepackage{soul}

\newcommand{\msun}{\mathrm{M}_\odot}
\graphicspath{{figures/}}

\def\lsim{ \lower .75ex \hbox{$\sim$} \llap{\raise .27ex \hbox{$<$}} }

\def\vxm{\overline{v_{\mathrm{x',i}}}}
\def\vym{\overline{v_{\mathrm{y',i}}}}
\def\vzm{\overline{v_{\mathrm{z',i}}}}

\def\vsqxm{\overline{v^2_{\mathrm{x',i}}}}
\def\vsqym{\overline{v^2_{\mathrm{y',i}}}}
\def\vsqzm{\overline{v^2_{\mathrm{z',i}}}}

\def\vsqxym{\overline{v^2_{\mathrm{x'y',i}}}}
\def\vsqxzm{\overline{v^2_{\mathrm{x'z',i}}}}
\def\vsqyzm{\overline{v^2_{\mathrm{y'z',i}}}}

\usepackage{lineno}
\submitjournal{ApJ}

\shorttitle{Testing \textsc{\textsc{jam}} with galaxy clusters in TNG300}
\shortauthors{Shi et al.}

\graphicspath{{./}{figures/}}

\begin{document}

\title{Inferring the mass content of galaxy clusters with satellite kinematics and Jeans Anisotropic modeling}

\correspondingauthor{Wenting Wang}
\email{wenting.wang@sjtu.edu.cn}

\author[0000-0001-7404-3706]{Rui Shi}
\affiliation{Department of Astronomy, Shanghai Jiao Tong University, Shanghai 200240, China}
\affiliation{Shanghai Key Laboratory for Particle Physics and Cosmology, Shanghai 200240, China}
\author[0000-0002-5762-7571]{Wenting Wang}
\affiliation{Department of Astronomy, Shanghai Jiao Tong University, Shanghai 200240, China}
\affiliation{Shanghai Key Laboratory for Particle Physics and Cosmology, Shanghai 200240, China}
\author[0000-0001-7890-4964]{Zhaozhou Li}
\affiliation{Centre for Astrophysics and Planetary Science, Racah Institute of Physics, The Hebrew University, Jerusalem, 91904, Israel}
\author{Ling Zhu}
\affiliation{Shanghai Astronomical Observatory, Chinese Academy of Sciences, 80 Nandan Road, Shanghai 200030, China}
\author[0000-0002-3712-6892]{Alexander Smith}
\affiliation{Institute for Computational Cosmology, Department of Physics, Durham University, South Road, Durham DH1 3LE, UK}
\author{Shaun Cole}
\affiliation{Institute for Computational Cosmology, Department of Physics, Durham University, South Road, Durham DH1 3LE, UK}
\author{Hongyu Gao}
\affiliation{Department of Astronomy, Shanghai Jiao Tong University, Shanghai 200240, China}
\affiliation{Shanghai Key Laboratory for Particle Physics and Cosmology, Shanghai 200240, China}
\author{Xiaokai Chen}
\affiliation{Department of Astronomy, Shanghai Jiao Tong University, Shanghai 200240, China}
\affiliation{Shanghai Key Laboratory for Particle Physics and Cosmology, Shanghai 200240, China}
\author{Qingyang Li}
\affiliation{Department of Astronomy, Shanghai Jiao Tong University, Shanghai 200240, China}
\affiliation{Shanghai Key Laboratory for Particle Physics and Cosmology, Shanghai 200240, China}
\author{Jiaxin Han}
\affiliation{Department of Astronomy, Shanghai Jiao Tong University, Shanghai 200240, China}
\affiliation{Shanghai Key Laboratory for Particle Physics and Cosmology, Shanghai 200240, China}

\begin{abstract}
Satellite galaxies can be used to indicate the dynamical mass of galaxy groups and clusters. In this study, we apply the axis-symmetric Jeans Anisotropic Multi-Gaussian Expansion (\textsc{JAM}) modeling to satellite galaxies in 28 galaxy clusters selected from the TNG300-1 simulation with halo mass of $\log_{10}M_\mathrm{200}/\msun>14.3$. If using true bound satellites as tracers, the best constrained total mass within the half-mass radius of satellites, $M(<r_\mathrm{half})$, and the virial mass, $M_{200}$, have average biases of -0.01 and $0.03$~dex, with average scatters of 0.11~dex and 0.15~dex. If selecting companions in redshift space with line-of-sight depth of 2,000~km/s, the biases are -0.06 and $0.01$~dex, while the scatters are 0.12 and 0.18~dex for $M(<r_\mathrm{half})$ and $M_{200}$. By comparing the best-fitting and actual density profiles, we find $\sim$29\% of best-fitting density profiles show very good agreement with the truth, $\sim$32\% display over or under estimates at most of the radial range with biased $M(<r_\mathrm{half})$, and 39\% show under/over estimates in central regions and over/under estimates in the outskirts, with good constraints on $M(<r_\mathrm{half})$, yet most of the best constraints are still consistent with the true profiles within 1-$\sigma$ statistical uncertainties for the three circumstances. Using a mock DESI Bright Galaxy Survey catalog with the effect of fiber incompleteness, we find DESI fiber assignments and the choice of flux limits barely modify the velocity dispersion profiles and are thus unlikely to affect the dynamical modeling outcomes. Our results show that with current and future deep spectroscopic surveys, \textsc{JAM} can be a powerful tool to constrain the underlying density profiles of individual massive galaxy clusters. 
\end{abstract}


\section{Introduction}
\label{sec:intro}

Galaxy clusters in our Universe, which contribute to the most luminous end of galaxy distribution and are hosted by the most massive populations of dark matter halos, are essential objects to study \citep[e.g.][]{Yang2007,Rykoff2014,Yang2021}. They provide suitable environments to examine the quenching of star formation in both the central massive galaxies and other smaller member satellite galaxies \citep[e.g.][]{Kimm2009,Wetzel2013,Boselli2016,Wang2018,PC2019}, to investigate the hot gas distribution through X-ray and Sunyaev-Zeldovich (SZ) observations \citep[e.g.][]{Arnaud2010,Planck2013,Lim2018}, to study the connection between galaxies, hot gas and the host dark matter halos \citep[e.g.][]{Planck2013,Anderson2015,Wang2016}, to look for missing baryons \citep[e.g.][]{Monteagudo2015,Graaff2019,Lim2020} and even serve as promising standard rulers in cosmology \citep[e.g.][]{Wagoner2021}. 

In the era of precision cosmology, accurate determination of the total mass of galaxy clusters, which is dominated by invisible dark matter, is a very important prerequisite for robust scientific conclusions in these different fields. Observationally, there are a few different approaches to constrain the mass of galaxy clusters. This includes non-kinematical methods of weak gravitational lensing \citep[e.g.][]{Rasia2012,Han15,Sun2021} and modeling of the redshift distortions \citep[e.g.][]{2012ApJ...758...50L}. Other kinematical methods include, for example, mass estimates based on the overall line-of-sight velocity (LOSV) dispersion of member satellite galaxies through calibrations with numerical simulations \citep[e.g.][]{2007MNRAS.379.1464S} and through the Halo Occupation Distribution framework \citep[e.g.][]{2009MNRAS.392..917M,2009MNRAS.392..801M,2011MNRAS.410..210M}, the caustic method \citep[e.g.][]{1997ApJ...481..633D,1999MNRAS.309..610D,2013ApJ...773..116G}, dynamical modeling of the observed hot gas distribution \citep[e.g.][]{Rasia2012,2017A&A...606A.122F}, virial theorem \citep[e.g.][]{Biviano2006}, Jeans or other more sophisticated dynamical modeling \citep[e.g.,][]{Mamon2013,Old2014} and machine learning \citep[e.g.][]{Ramanah2021} approaches to recover the cluster mass from the projected phase-space distribution of satellite galaxies, and more recently, a combination of the satellite kinematics and luminosity functions under a hierarchical Bayesian inference formalism \citep{2019MNRAS.488.4984V}.

Among the different satellite kinematical based methods above, the virial mass estimator and the machine learning approach usually give a single estimate of the total cluster mass. The machine learning approach, the empirical relation deduced by \cite{2007MNRAS.379.1464S} and the modeling of redshift distortion \citep{2012ApJ...758...50L} often rely on external numerical simulations. Compared with the other methods, dynamical modeling of satellite galaxies can in principle constrain a parameterized mass or potential model, and does not require external simulations, but it requires a relatively large sample of satellite galaxies as dynamical tracers. 

There are many deep spectroscopic surveys, such as the Sloan Digital Sky Survey-V
\citep[SDSS-V;][]{2017arXiv171103234K}, the Subaru Prime Focus Spectroscopy \citep[PFS;][]{2014PASJ...66R...1T}, the Dark Energy Spectroscopic Instrument \citep[DESI;][]{2016arXiv161100036D,2023AJ....165...50M}, and future Stage-5 spectroscopic instruments such as MegaMapper \citep{2022arXiv220904322S}, the Mauna Kea Spectroscopic Explorer (MSE). For the most massive galaxy clusters in our local Universe, it is very promising to obtain the line-of-sight velocities for $>\sim$100 member satellite galaxies, hence enabling the modeling of the parameterized mass/potential profiles, instead of only a single value of the total mass. 

However, like the virial theorem, dynamical equilibrium has to be assumed for almost all dynamical modeling approaches. Since massive galaxy clusters assemble late, they may deviate more from equilibrium than less massive galaxy groups. The modeling outcome may be biased from the truth. In order to understand the amount of biases upon the dynamical modeling of galaxy clusters at first, we adopt realistic galaxy cluster systems from the Illustris-TNG300 simulations \citep{Springel2018} to test the model performance with the axis-symmetric Jeans Anisotropic Multi-Gaussian Expansion modeling method \citep[\textsc{jam;}][]{Cappellari2008,Watkins2013}. In our work, we directly know the true density profiles from the simulations, and we apply \textsc{jam} to the kinematics of simulated satellite galaxies, to recover the mass density profiles. In this way, we are capable of evaluating the model performances and biases, before applying the method to real data in our planned future studies. In addition to the Illustris-TNG300 simulation, we also adopt a mock DESI bright galaxy survey (BGS) catalog \citep{2017MNRAS.470.4646S}, to investigate observational effects including the fiber incompleteness and the dependence on the survey flux limit. 

The layout of this paper is as follows. We introduce the TNG suites of simulations, our selections of galaxy clusters, satellite galaxies as tracers and the creation of mock galaxy images and multi-Gaussian expansion in Section~\ref{sec:data}. Section~\ref{sec:methods} provides an introduction to the dynamical modeling method. Results will be presented in Section~\ref{sec:results}, including demonstrations of the model performance based on bound satellites and satellites selected in redshift space with contaminations. We also discuss the impact of fiber incompleteness and the effect of flux limits on our analysis. We conclude in Section~\ref{sec:concl}.

\section{Data}
\label{sec:data}

\subsection{The IllustrisTNG simulation}
\label{sec:TNG300}

The sample of galaxy clusters is constructed from the TNG300-1 simulation of the IllustrisTNG Project \citep{Pillepich2018,Springel2018}. The IllustrisTNG simulations are a suite of hydrodynamical simulations incorporating sophisticated baryonic processes, carried out with a moving-mesh code \citep[\textsc{arepo};][]{Springel2010} to solve the equations of gravity and magneto-hydrodynamics. They include comprehensive treatments of various galaxy formation and evolution processes, such as metal line cooling, star formation and evolution, chemical enrichment and gas recycling. For more details about TNG, we refer readers to \cite{Marinacci2018,Naiman2018,Nelson2018,Nelson2019}.

The TNG300 suite of simulations adopt the Planck 2015 $\Lambda$CDM cosmological model with $\Omega_\mathrm{m}=0.3089$, $\Omega_\Lambda=0.6911$, $\Omega_\mathrm{b}=0.0486$, $\sigma_8=0.8159$, $n_s=0.9667$, and $h=0.6774$ \citep{Planck2015}. TNG300-1 is the simulation with the highest resolution in its suite (compared with TNG300-2 and TNG300-3), and hereafter we refer to it as TNG300. It has a periodic comoving box with 302.6~Mpc on each side that follows the joint evolution of 2,500$^3$ dark matter particles and approximately 2,500$^3$ baryonic resolution elements (gas cells and star particles). Each dark matter particle has a mass of $5.9\times10^7 \rm{M}_\odot$, while the baryonic mass resolution is $1.1\times10^7 \msun$. Collisionless particles, such as dark matter and stars, have a softening length of 1.5~kpc whereas gas particles have variable softening scales with a minimum of 370~pc.

\subsection{Galaxy cluster systems in TNG}
\label{sec:galaxy}

\begin{figure*}
\includegraphics[width=1.0\textwidth]{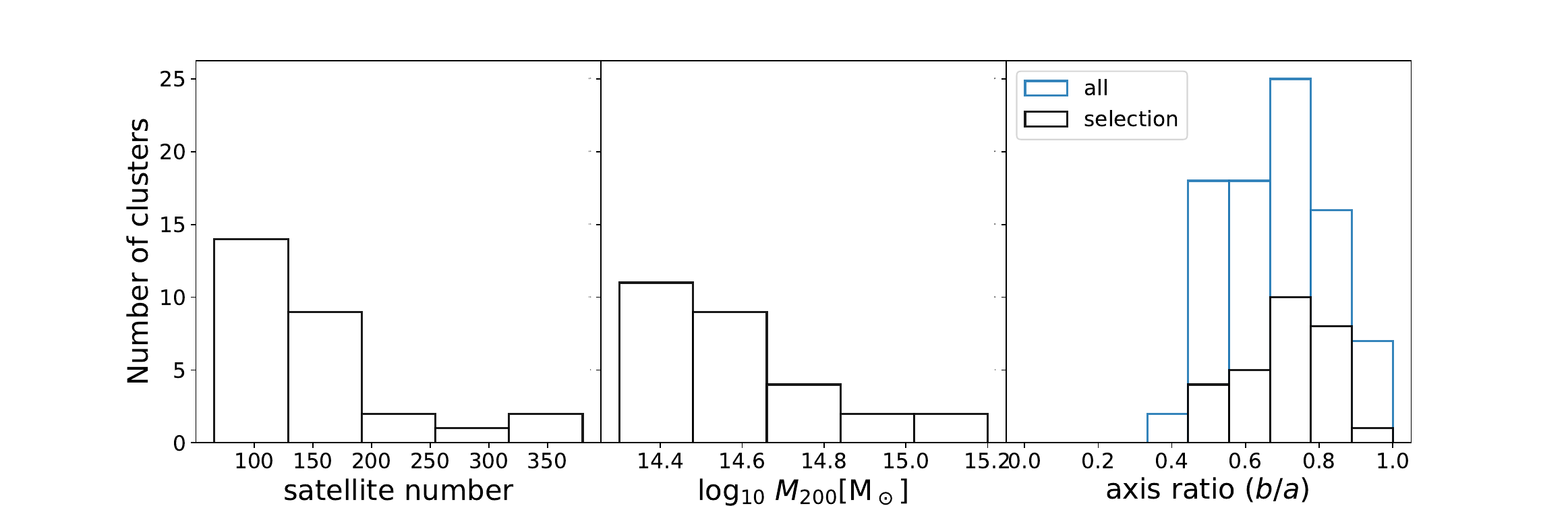}%
\caption{The distribution of the number of bound satellites in each cluster (left), the virial mass ($M_{200}$, middle) and the middle to major axis ratio ($b/a$, right) for our mock galaxy clusters from TNG300. In the right panel, the black and blue histograms refer to our selected galaxy clusters and all clusters with the same halo mass threshold in TNG300, respectively.}
\label{fig:frequency}
\end{figure*}

Dark matter halos in TNG are identified with the friends-of-friends (FoF) algorithm \citep{Davis1985}. In each FoF group, substructures (subhalos/galaxies) are identified with the SUBFIND algorithm~\citep{SUBFIND}. The most massive subhalo in each group, together with its baryonic component, is called the main subhalo and the central galaxy. All other subhalos/galaxies in the halo are referred to as satellites. 

In our study, we first select massive galaxy clusters with $M_{200}>10^{14.3}\msun$\footnote{The virial mass, $M_{200}$, is defined as the mass enclosed in a radius, $R_{200}$, within which the mean matter density is 200 times the critical density of the universe.} from the redshift $0$ snapshot of TNG300. There are 86 galaxy clusters falling within this mass range in TNG300. Most of these massive clusters with $M_{200}>10^{14.3}\msun$ can have more than 100 bound satellites\footnote{Bound satellites are defined as those companion galaxies around each galaxy cluster system that have total energy smaller than zero in the simulation.} with a stellar mass threshold of $M_\odot>10^{9}\msun$\footnote{This mass threshold is chosen to ensure that the satellites can have more than $\sim$100 star particles in TNG300.} and projected within 2~Mpc. At the lower boundary of $M_{200}\sim 10^{14.3}\msun$, the minimal number of bound satellites is $\sim$70. The number of satellites is enough for dynamical modeling. We further select cluster systems by requiring the central galaxies of these clusters to be at least 2 magnitudes brighter in $r$-band than the brightest companions projected within 4~Mpc and with the line-of-sight velocity differences with respect to the central galaxies smaller than 2,000~km/s. These selections result in 28 clusters that meet these requirements.

In Figure~\ref{fig:frequency}, we show the distribution of the number of bound satellites and $M_{200}$ for these galaxy clusters. In particular, we show in the right panel of Figure~\ref{fig:frequency} the distribution of middle to major axis ratios ($b/a$) for our selected galaxy cluster systems (black) and all galaxy clusters with the same mass threshold of $M_{200}>10^{14.3}\msun$ in TNG300. Here $b/a$ is calculated by using all bound star particles within $R_{200}$ in these cluster systems, including those in bound satellite galaxies. Our selected clusters have $b/a>0.45$.%, which have a slight bias towards larger $b/a$ values compared to all clusters. This is mainly due to the elimination of massive companions in our selections.

We choose the $Z$-axis\footnote{In this paper, we will have three different coordinate systems. The first one is the $X$, $Y$ and $Z$-axes of the simulation box, which we denote using capital letters. The observing frame is defined using letters with a prime symbol, i.e., $x'$, $y'$ and $z'$. In Section~\ref{sec:jeans} below, we will define another intrinsic coordinate system centered on the central galaxy, which we denote using $x$, $y$ and $z$.} of the TNG300 simulation box as the line-of-sight direction. We define, for the observing frame, the $z'$-axis as aligned with the line-of-sight direction, and the $x'-y'$ plane to be perpendicular to the line-of-sight direction. Here the $x'$-axis is defined as the image major axis of the galaxy cluster in projection. Notably, the observing frame is a left-handed system. 

The central coordinate of each galaxy cluster is defined as the potential minimum of the main subhalo, and the velocity of each cluster is defined as the mass-weighted and averaged velocity based on all particles in the main subhalo. The velocity of each satellite is calculated relative to the velocity of the cluster after considering the Hubble flow. Explicitly, the line-of-sight velocities of satellites in galaxy clusters from TNG300 are calculated as $v_\mathrm{los}=H_0(Z-Z_\mathrm{cen})+(v_Z-v_{Z,\mathrm{cen}})$. Here $Z_\mathrm{cen}$ and $v_{Z,\mathrm{cen}}$ are the $Z$ coordinate and the velocity along the $Z$-axis of the simulation box for the cluster center, while $Z$ and $v_Z$ are the corresponding coordinate and velocity for the satellite. $H_0$ is the Hubble constant at redshift 0.

For results based on TNG300 in this study, we select dynamical tracer satellite galaxies in two different ways. We first select only true bound satellite galaxies projected within 2~Mpc and more massive than $10^9\msun$ as dynamical tracers. Then, to mimic real observation, satellites are selected as those projected within 2~Mpc, within 2,000~km/s along the line-of-sight direction and also more massive than $10^9\msun$. Our choice of the line-of-sight depth is based on a natural boundary of dark matter halos revealed around the minimum bias and maximum infall locations, and this boundary is very close to twice the virial radius of dark matter halos, which is close to 2,000~km/s along the line of sight for our massive galaxy cluster systems \citep[e.g.][]{2021MNRAS.503.4250F,2022MNRAS.513.4754F,2023ApJ...953...37G}. We find the completeness of satellites selected in this way is $\sim$88\% on average, and the contamination is $\sim$11\%. In our analysis, we will test how the dynamical modeling outcome changes with the contamination. 

\subsection{The DESI BGS mock catalog}
\label{sec:mxxl}

In addition to TNG galaxy cluster systems, we use a mock DESI Bright Galaxy Survey (BGS) catalog \citep{2017MNRAS.470.4646S} to investigate observational effects including fiber incompleteness and flux limits. The DESI BGS survey is expected to cover an area of $\sim$14,000 square degrees in 4 passes of the sky, with a depth approximately two magnitudes deeper than that of the SDSS, hence providing more spectroscopically observed satellite galaxies in galaxy clusters \citep{2023AJ....165..253H}. 

The BGS mock catalog we adopted in this study is based on the Millennium-XXL (MXXL) simulation \citep{2012MNRAS.426.2046A}, which adopts the WMAP1 cosmological parameters of $\Omega_\mathrm{m}=0.25$, $\Omega_\Lambda=0.75$,  $\sigma_8=0.9$, $n=1$, and $h=0.73$. It is a light-cone mock catalog \citep{2017MNRAS.470.4646S}, which covers the full sky and extends to redshift $0.8$ with a mass resolution of $\sim10^{11.14}\mathrm{M_\odot}$. The lightcone is created with interpolation between different snapshots \citep{2013MNRAS.429..556M}.

Satellites are randomly positioned following a Navarro-Frenk-White (NFW) \citep{1997ApJ...490..493N,1996ApJ...462..563N} density profile, with randomly assigned velocities following the Maxwell–Boltzmann distribution. A Monte Carlo method is used to assign a $r$-band magnitude and a $g-r$ color to each galaxy to build a galaxy catalog whose luminosity function of galaxies is in agreement with Sloan Digital Sky Survey \citep[SDSS;][]{2009ApJS..182..543A} and the Galaxy and Mass Assembly survey \citep[GAMA;][]{2009A&G....50e..12D,2011MNRAS.413..971D}. The galaxy catalog has a flux limit of $r<20$ and a median redshift of $z\sim0.2$. The flux limit of $r<20$ is faint enough for the DESI BGS bright sample, as the BGS bright sample has a flux limit of $r<19.5$. The fiber assignment algorithm has been run on the mock \citep{2019MNRAS.484.1285S}, enabling us to quantify the impact of fiber assignment in galaxy surveys. 

Note, however, the fiber assignment algorithm of \cite{2019MNRAS.484.1285S} is currently being updated with the progress of the DESI observation. The mock BGS catalog we are currently using in this study \citep{2017MNRAS.470.4646S} is based on three passes, and it is now being updated to four passes. Moreover, the DESI BGS survey has a faint sample down to a flux limit of $r<20.175$, in order to increase the overall BGS target density and enable small-scale clustering measurements \citep{2023AJ....165..253H}. Thus the flux limit is being updated from $r<20$ to $r<20.175$ in the latest DESI BGC mock under construction. Nevertheless, we think these improvements and modifications will not affect our conclusions (see Section~\ref{sec:fiber incompleteness} for details).

We select tracer satellites from the DESI BGS mock as those companions that are projected within $R_{200}$ and with line-of-sight velocity differences with respect to the central galaxy smaller than twice the virial velocity of the host halos. After applying DESI masks, we find that there are 84 galaxy clusters with redshifts lower than 0.2 within the DESI Year 5 footprint, which can have more than 100 satellite galaxies with $r<19.5$ selected in this way. The line-of-sight velocities ($v_{\rm los}$) of the satellite galaxies in the BGS mock catalog are calculated based on the following equation
\begin{equation}
    v_\mathrm{los}=\frac{c(\mathrm{redshift}-\mathrm{redshift}_\mathrm{cen})}{1+\mathrm{redshift}_\mathrm{cen}},
\end{equation}
where $c$ is the speed of light, $\mathrm{redshift}-\mathrm{redshift}_\mathrm{cen}$ is the difference between the redshifts of the satellite galaxy and the central galaxy of the galaxy cluster. 

The mock DESI BGS catalog is based on different cosmological parameters from those of TNG, with satellites populated in dark matter halos in different ways. In principle, it is better to focus on the same simulation, but TNG does not have realistic light-cone mocks. We thus checked the average distribution of line-of-sight velocity, velocity dispersion and surface number density profiles of satellite galaxies in the two simulated data sets (TNG and DESI BGS catalog), and we find consistent spatial and velocity distributions of satellites, which ensures a fair usage of the mock DESI BGS catalog.

\subsection{Mock galaxy cluster images and multi-Gaussian decomposition}
\label{sec:MGE}

\begin{figure*} 
\includegraphics[width=1\textwidth]{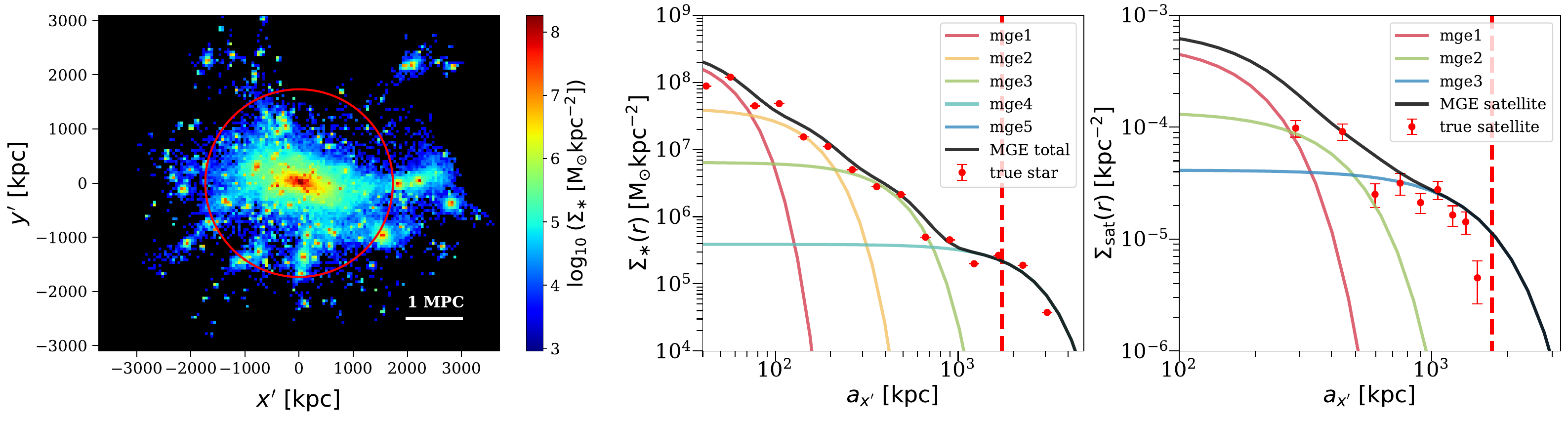}%
\caption{{\bf Left:} The surface density map for the stellar component of one representative galaxy cluster system from TNG300. The red circle corresponds to $R_{200}$. {\bf Middle:} Red dots with errorbars are the true surface density profile of the total stellar mass distribution. This is calculated from elliptical isophotes and reported as a function of the major axis length ($a_{x'}$) for the same representative galaxy cluster in the left plot. The errorbars are the 1-$\sigma$ scatters based on 100 bootstrap samples of all star particles in TNG300, which are comparable to the symbol size. We show each individual MGE component with a different colored curve, and the black solid curve is the total best-fitting surface density profile, contributed by the combination of all MGEs. The red vertical dashed line marks $R_{200}$. {\bf Right:} Similar to the left panel, but comparing the true projected satellite number density profile and the MGE decompositions. The errorbars are the 1-$\sigma$ scatters among 100 bootstrap satellite samples.}
\label{fig:mge}
\end{figure*}

In our study, we aim to constrain the underlying potential using satellite galaxies as dynamical tracers. The potential is contributed by both luminous and dark matter. \textsc{jam} directly infers the potential formed by the luminous matter distribution from the deprojected ``optical'' image. Hence we need to create mock galaxy cluster images for our analysis. For the mock images of galaxy clusters, we simply adopt the projected stellar mass density or surface density to create the images, i.e., the reading in each pixel is in units of $\msun/\mathrm{pc}^2$ based on all bound star particles associated with the galaxy cluster. Note the mock images of galaxy clusters are contributed by all star particles bound to the cluster and also those star particles in satellite galaxies bound to the cluster. In real observations, however, the observed diffuse light in the outskirts of the clusters depends on the surface brightness limit of the optical survey, but can be well measured for individual galaxy clusters in modern deep photometric surveys \citep[e.g.][]{2018MNRAS.475.3348H,2019MNRAS.487.1580W}.

The projected number density distribution of tracer satellites plays a critical role in solving the Jeans equation (see Section~\ref{sec:jeans} for more details). Therefore, we also create the projected satellite number density maps for each galaxy cluster system, based on the projected positions of selected tracer satellite galaxies, with each satellite contributing the same weight, regardless of its actual stellar mass or luminosity in the simulation.

Once the mock images or maps are made, the projected luminous stellar mass distributions and the projected satellite number distributions will be decomposed to Multiple Gaussian Elements \citep[MGE;][]{Emsellem1994,DSouza2013}, in order to enable the analytical deprojection for each MGE component to 3-dimensions and to bring analytical solutions for any arbitrary matter distribution (see Section~\ref{sec:methods} for more details). 

In practice, we execute the MGE decomposition with the sherpa software \citep{sherpa1,sherpa2}, which is a modeling and fitting module integrated with CIAO to fit the mock images and optimize the solutions of each MGE. 

The surface density distribution is shown for one example galaxy cluster from TNG300 in the left plot of Figure~\ref{fig:mge}. Note \textsc{jam} requires the major axes of the mock galaxy and tracer images to align with the $x'$ axes of the image plane, and thus the galaxy cluster has been rotated to meet the requirement. 

In the middle panel, we show each individual Gaussian element of the stellar component with different colored curves for this cluster. The combined surface density profile of all MGEs is represented by the black curve, and the true surface density profile of the stellar component is shown by red dots. From this plot, we can see the red dots agree well with the black line from $\sim$100 kpc to $R_{200}$, indicating a good overall performance of the MGE decomposition. In the inner-most region (<100kpc), the black line tends to be slightly higher than the red dots, and such a bias is primarily caused by the lower number of pixels in such central regions. The right panel of Figure~\ref{fig:mge} is similar to the left one, but it shows the MGE decomposition for the projected satellite/tracer number density profile. The errorbars of the red dots are based on 100 bootstrap samples of satellites in projection. In detail, we perform multiple bootstrap samples by randomly selecting a subset of these satellites each time with repeats and calculating the 1-$\sigma$ scatters of the projected satellite number density profiles. Because of the limited number of satellites, the errorbars are significantly larger in the right plot, and the MGE decomposition is not as good as in the middle panel, especially in the very inner region and the outer region close to $R_{200}$.

\section{Methodology}
\label{sec:methods}

\subsection{The Jeans equation and potential model}
\label{sec:jeans}

Jeans Anisotropic Multi-Gaussian Expansion (\textsc{jam}) modeling is a powerful dynamical modeling tool, which can be used to constrain the luminous and dark matter distributions of globular clusters, dwarf galaxies and distant galaxies with Integral Field Unit (IFU) observations. In this study, we investigate the performance of \textsc{jam} when it is applied to galaxy clusters, with satellite galaxies in the cluster as dynamical tracers. A detailed description of \textsc{jam} can be found in \cite{Cappellari2008} and \cite{Watkins2013}. Here we only give a brief introduction. 

The Jeans equation for an axis-symmetric system ($\frac{\partial}{\partial \phi}=0$) in steady state ($\frac{\partial}{\partial t}=0$) can be written in cylindrical coordinates as:
\begin{equation}
    \frac{\nu(\overline{v_R^2}-\overline{v_\phi^2})}{R}+\frac{\partial(\nu \overline{v_R^2})}{\partial R}+\frac{\partial (\nu \overline{v_R v_z})}{\partial z}=-\nu \frac{\partial \Phi_{\rm{tot}}}{\partial R},
    \label{eqn:jeans1}
\end{equation}
\begin{equation}
    \frac{\nu \overline{v_R v_z}}{R}+\frac{\partial (\nu \overline{v_R v_z})}{\partial R}+\frac{\partial (\nu \overline{v_z^2})}{\partial z}=-\nu \frac{\partial \Phi_{\rm{tot}}}{\partial z},
    \label{eqn:jeans2}
\end{equation}
where $\nu$ is the satellite number density distribution, which has been decomposed into a few different MGEs (see Section~\ref{sec:MGE} for details). $\Phi_{\rm{tot}}$ is the total gravitational potential.

\textsc{jam} models the total potential, $\Phi_{\rm{tot}}$, with two different components: 1) the stellar component, as we have mentioned in Section~\ref{sec:MGE}, is deprojected and evaluated from the surface density distribution\footnote{The stellar component is subdominant compared with dark matter in galaxy clusters, and throughout this paper, we focus our discussions on the constraints of the total matter distribution.}; and 2) the dark matter component\footnote{\textsc{jam} does not model the gas component separately, because in real observation, the spatial distribution of hot and cold gas in distant galaxies is often difficult to be directly observed with high resolution. In principle, we can modify \textsc{jam} to model the gas component separately. However, we have checked that the hot gas component is distributed over the whole halo and has similar radial distributions as that of dark matter. Even if we model the gas and dark matter components separately, \textsc{jam} would fail to distinguish them. So for our analysis throughout this paper, the gas component would be modeled within the dark matter component. Since our main conclusions are based on the total matter distribution, whether the gas component is modeled separately or not is not important.}. To model the dark matter component, we simply adopt the following double power-law model:

\begin{equation}
    \rho(r)=\frac{\rho_s}{(r/r_s)^\gamma (1 +r/r_s)^\alpha},
    \label{eqn:double}
\end{equation}
in which $\rho_s$ and $r_s$ are the scale density and scale radius, and $\alpha$ is the outer power law index. In this study, due to the significantly lower number of satellite galaxies in central regions of the galaxy clusters, we find our constraints on the inner density slope, $\gamma$, is very weak, so we fix $\gamma$ to be 1 throughout this paper. 

As we have mentioned in Section~\ref{sec:MGE}, \textsc{jam} determines the potential contributed by the stellar mass distribution from the deprojected optical images of galaxy clusters. Since the pixel units of our mock images are $M_\odot$/pc$^2$, we simply fix the stellar-mass-to-light ratio ($M_\ast/L$) to unity in our analysis. We decompose the stellar mass distribution into MGEs. Each component enables fast deprojections and leads to quick analytical solutions to the Jeans equation. 
Given a model dark matter density profile, we also decompose it into a few different MGEs (see Section~\ref{sec:MGE} for details). In this way, we have MGE components for the luminous and dark matter potential, and tracer number density distribution, $\nu$. We can thus have analytical solutions for each MGE component. The final solution to the above Jeans equation (Equations~\ref{eqn:jeans1} and \ref{eqn:jeans2}) is the summation of all different components. 

To ensure the Jeans equation having unique solutions of the first and second velocity moments, the velocity ellipsoid is further assumed to be aligned with the cylindrical polar coordinate system ($\overline{v_R v_z}=0$). In addition, a constant anisotropy parameter, $\lambda$, is introduced as $\overline{v_R^2}=\lambda\overline{v_z^2}$. A rotation parameter, $\kappa$, is introduced as $\overline{v_\phi}=\kappa
(\overline{v_\phi^2}-\overline{v_R^2})^{1/2}$, with the calculation of it modified according to \cite{Zhu2016}, though for galaxy clusters, which are not rotation dominated systems, $\kappa$ is not expected to be significantly different from zero. In principle, $\kappa$ can be either positive or negative, depending on the direction of rotation, i.e., clockwise or counter-clockwise seeing from the positive $z$-axis in the intrinsic frame (see the definition below for the intrinsic frame). With the boundary condition set to $\nu \overline{v_z^2}=0$ as $z \rightarrow \infty$, the Jeans equation can be summarised as:
\begin{equation}
\nu \overline{v_\phi^2}(R, z)=\lambda\left[R \frac{\partial\left(\nu \overline{v_R^2}\right)}{\partial R}+\nu \overline{v_z^2}\right]+R \nu \frac{\partial \Phi}{\partial R}
\end{equation}
\begin{equation}
\nu \overline{v_z^2}(R, z)=\int_z^{\infty} v \frac{\partial \Phi}{\partial z} \mathrm{d}z.
\end{equation}

Given a model potential, the velocity first and second moments are solved in an intrinsic frame defined on the cluster system. The intrinsic frame is a right-handed system, and we use the quantities without the prime to denote the intrinsic frame, i.e., $x$, $y$ and $z$. In our analysis the $z$-axis is chosen to be the minor axis of the cluster system, with the minor axis calculated from the spatial distribution of satellites. The $x$-axis is chosen to be aligned with the $x'$-axis of the observing frame (see Section~\ref{sec:galaxy} above), with the $x'$-axis being the projected major axis of the galaxy cluster in the image plane. $y$-axis is determined according to the $x$-axis and $z$-axis to form the right-handed cartesian coordinate system. The intrinsic frame can be linked to the left-handed observing frame (see Section~\ref{sec:galaxy}) through the following equations
\begin{equation}
    \left( \begin{array}{c}
        x' \\
        y' \\
        z'
    \end{array} \right) = \left( \begin{array}{ccc}
        1 & 0 & 0 \\
        0 & - \cos(incl) & \sin(incl) \\
        0 & \sin(incl) & \cos(incl)
    \end{array} \right) \left( \begin{array}{c}
        x \\
        y \\
        z
    \end{array} \right),
\end{equation}
where $incl$ is the inclination angle, defined as the angle between the $z$-axis of the intrinsic frame and the $z'$-axis of the observing frame.

The first and second moments of line-of-sight velocities, solved from the Jeans equation mentioned above, can be compared with the actual velocity moments of tracer satellites. Then the best potential model parameters can be inferred by maximizing the likelihood function, which will be introduced in the next section.

\subsection{Likelihood function}

We model the posterior probability distribution of our model parameters by Bayes theorem

\begin{equation}
p(\boldsymbol{\Theta} \mid \boldsymbol{D})=\frac{p(\boldsymbol{D} \mid \boldsymbol{\Theta}) p(\boldsymbol{\Theta})}{p(\boldsymbol{D})}.
\label{eq:bayes}
\end{equation}

Our list of model parameters is $\boldsymbol{\Theta}=(\lambda,\kappa,\rho_s,r_s,\alpha)$ (see Section~\ref{sec:jeans} above or a summary of free parameters near the end of this subsection). 
$p(\boldsymbol{\Theta})$ is the prior, and $p(\boldsymbol{D|\Theta})$ is the distribution of the velocities by assuming that the prediction of the velocity distribution obeys the multivariate Gaussian distribution. $p(\boldsymbol{D})$ is a factor required to normalize the posterior.

The likelihood of the satellites in each cluster system can be written as:
\begin{equation}
    \begin{split}
    L_{\mathrm{tot}}^{\mathrm{sat}} &=p(\boldsymbol{D} \mid \boldsymbol{\Theta}) \\
    &=\prod_{i=1}^{N_\mathrm{sat}}{p\left(\boldsymbol{v}_i \mid \boldsymbol{x}_i^{\prime}, \boldsymbol{S}_i, \Theta\right)}\\
    &= \prod_{i=1}^{N_\mathrm{sat}} \frac{\exp \left[-\frac{1}{2}\left(\boldsymbol{v}_i-\boldsymbol{\mu}_i\right)^{\mathrm{T}}\left(\boldsymbol{C}_i+\boldsymbol{S}_i\right)^{-1}\left(\boldsymbol{v}_i-\boldsymbol{\mu}_i\right)\right]}{\sqrt{(2 \pi)^3\left|\left(\boldsymbol{C}_i+\boldsymbol{S}_i\right)\right|}},
    \label{eqn:likelihood}
    \end{split}
\end{equation}
where $\boldsymbol{v}_i$ represents the velocity solved by \textsc{jam} at the position of the observed tracer satellite, $\boldsymbol{x'}_i=(x'_i,y'_i)$, and $\boldsymbol{\mu}_i$ is the observed volocity of the tracer satellite.

The covariance matrix $\boldsymbol{C}_i$ is defined through the first and second velocity moments:
 
\begin{align}
   & \boldsymbol{C}_i = \nonumber\\
   & \left( \begin{array}{ccc}
        \vsqxm - \vxm^2 & \vsqxym - \vxm\,\vym & \vsqxzm - \vxm\,\vzm \\
        \vsqxym - \vxm\,\vym & \vsqym - \vym^2 & \vsqyzm - \vym\,\vzm \\
        \vsqxzm - \vxm\,\vzm & \vsqyzm - \vym\,\vzm & \vsqzm - \vzm^2
    \end{array} \right),
\end{align}
and $\boldsymbol{S_i}$ is the error matrix of the observed velocity of a tracer satellite
\begin{equation}
  \boldsymbol{S}_i =
  \left( {\begin{array}{ccc}
    \sigma^2_{v_{x'},i} & 0 & 0 \\
    0 & \sigma^2_{v_{y'},i} & 0 \\
    0 & 0 & \sigma^2_{v_{z'},i} \\
  \end{array} } \right).
\end{equation}

Notably, in our work, we only use the line-of-sight velocities, so we simply set $v_{x',i}=v_{y',i}=0$ and input very large values for $\sigma^2_{v_{y'},i}$ and  $\sigma^2_{v_{z'},i}$. This is equivalent to only fitting the observed first and second moments of LOSVs.

\begin{equation}
m_i\left(\boldsymbol{x}_i^{\prime}\right)=\frac{\Sigma\left(\boldsymbol{x}_i^{\prime}\right)}{\Sigma\left(\boldsymbol{x}_i^{\prime}\right)+\epsilon \Sigma(0,0)}
\label{member prior}
\end{equation}
where $\Sigma(\boldsymbol{x}_i^{\prime})$ is the surface density at $\boldsymbol{x}_i^{\prime}$ and $\Sigma(0,0)$ is the central surface density. 

The likelihood of the fore/background satellites, $L_i^{\mathrm{bkgd}}$, can be calculated by assuming a tri-variate Gaussian distribution with a given mean velocity and velocity dispersion of a fore/background model. In our case, the mean velocity and velocity dispersion of the fore/background model are directly calculated from unbound satellite galaxies in the simulation. Then the likelihood becomes:
\begin{equation}
L=\prod_{i=1}^{N_\mathrm{sat}}m_i(\boldsymbol{x}_i^{\prime}) L_i^{\mathrm{sat}}+[1-m_i(\boldsymbol{x}_i^{\prime})] L_i^{\mathrm{bkgd}},
\label{likelohood_contamination}
\end{equation}

We summarize the list of our model parameters as follows:\\
(1) $\lambda$, the velocity anisotropy; \\
(2) $\kappa$, the rotation parameter;\\
(3) $d_1\equiv \mathrm{log_{10}}(\rho_s^2r_s^3)$;\\
(4) $d_2\equiv\mathrm{log_{10}}(\rho_s)$;\\
(5) $\alpha$, the outer density slope;\\
(6) $\epsilon$, the background fraction;\\
(7) $incl$, the inclination angle.

Here $d_1$ and $d_2$ are constructed to reduce the strong degeneracy between $\rho_s$ and $r_s$. The logarithmic transformation converts the original units of $\rm{M_{\odot}}^2/\mathrm{pc}^3$ and $\rm{M_{\odot}}/\mathrm{pc}^3$ into dimensionless logarithmic values. After taking the logarithm, they cover a smaller range in log space. $\epsilon$ is fixed to zero when we only consider bound satellites as tracers. To obtain the best-fitting model parameters,  we set a flat prior of $\boldsymbol{\Theta}$ and use the Markov chain Monte Carlo (MCMC) approach\footnote{When fitting the double power law function to the true density profiles, we use \textsc{emcee} \citep{Foreman-Mackey2013} to sample the posterior distribution of parameters.} to maximize the likelihood function.

In our analysis, we will try two different cases by either fixing $incl$ or allowing it to be a free model parameter. When fixing $incl$, its value is chosen as the angle between the line-of-sight direction of the mock observer and the minor axis calculated from the spatial distribution of bound satellites. However, since our galaxy cluster systems are not rotationally dominated and the minor axes of realistic galaxy cluster systems differ for different components, the definition of $incl$ is not straightforward. For example, if we calculate the minor axis according to the spatial distribution of the central galaxy, the angle between this minor axis and the line-of-sight direction would be different. For individual systems, fixing $incl$ according to different minor axes definitions or treating $incl$ as a free parameter can lead to different constraints, though for most of the time, they are still consistent within 1-$\sigma$ due to our small number of tracer satellites. For all 28 cluster systems, we will show later that the amounts of overall biases and scatters of either fixing $incl$ or treating it as a free parameter do not show significant differences. Setting $incl$ as a free parameter leads to slightly smaller scatters. Note in real observation, the constraint on $incl$ for galaxy cluster systems is weak and $incl$ is chosen to be fixed \citep[e.g.][]{2020MNRAS.492.2775L}.

We fix the stellar mass-to-light ratio, $M_\ast/L$, to their true values in the simulation. In our analysis, the true value of $M_\ast/L$ is unity. We do not test the uncertainties in $M_\ast/L$ in this study. In real observations, $M_\ast/L$ can be determined through stellar population synthesis modeling and fixed upon dynamical modeling. The uncertainties of $M_\ast/L$ by population synthesis modeling, however, depend on many different factors, including the number of available bands adopted for the synthesis modeling, the adopted stellar libraries, initial mass functions, dust models and so on\citep[e.g.][]{2013ARA&A..51..393C}, which might have significant systematic uncertainties but these are hard to be directly tested for real galaxies.

Nevertheless, we find the stellar component is more subdominant than the dark matter component for our galaxy cluster systems from TNG. Moreover, throughout our analysis in this paper, we focus on discussing the total mass profiles, so we believe uncertainties in $M_\ast/L$ would not significantly affect the generality of the main conclusions about the total mass profile in this paper. In fact, in a previous study, \cite{2022ApJ...941..108W} found that if $M_\ast/L$ is fixed to a significantly high value, \textsc{jam} would decrease the contribution by dark matter, which maintains almost the same best constrained total matter distribution. In principle, we can modify \textsc{jam} to let it directly model the total matter distribution for our sample of galaxy clusters, instead of modeling the stellar and dark matter components separately, but this is not incorporated for our analysis in this current paper.

\section{Results}
\label{sec:results}

In this subsection, we investigate the accuracy and bias of the mass profiles predicted by \textsc{jam}.

\subsection{The overall performance with massive galaxy clusters from TNG300}
\label{sec:overall}

\begin{figure*} 
\includegraphics[width=0.8\textwidth]{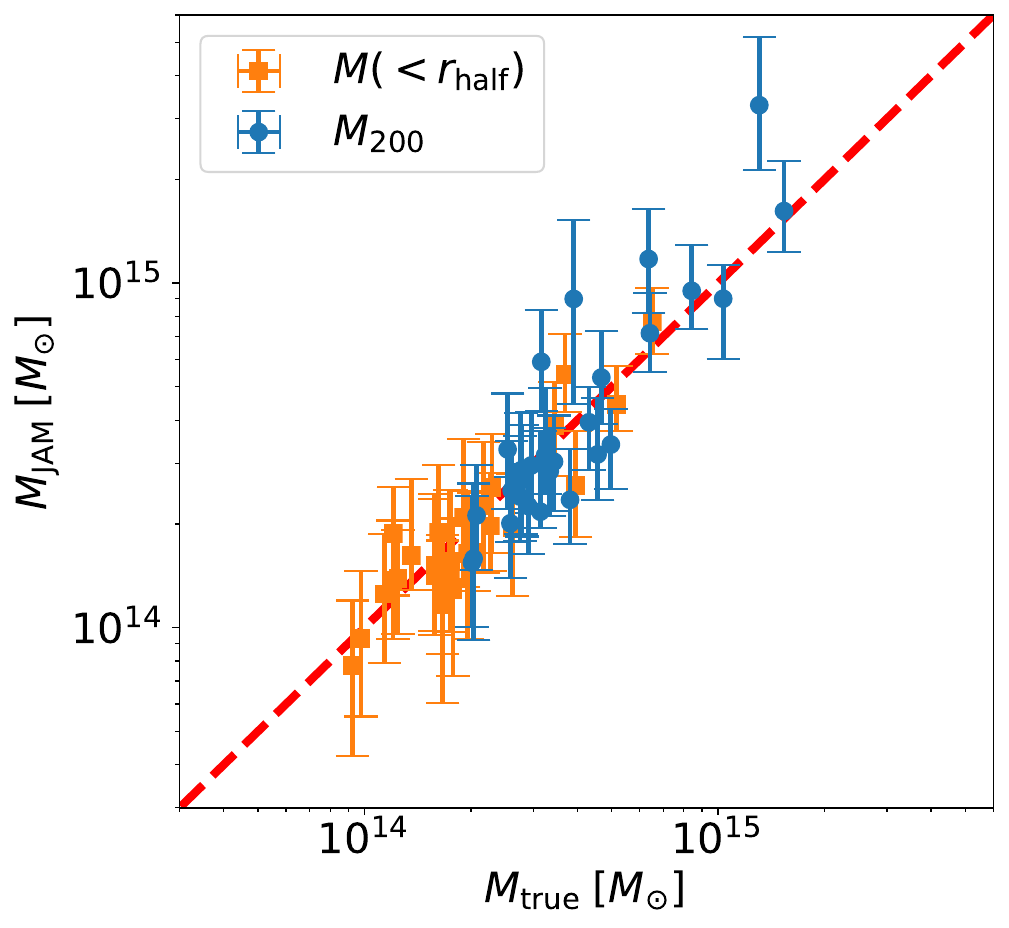}%
\caption{The blue dots with errorbars show the best-fitting $M_{200}$ by \textsc{jam} versus the true masses for 28 galaxy clusters from TNG300. Here we adopt true bound satellites as tracers, and the inclination, $incl$, is fixed to the angle between the line-of-sight direction of the mock observer and the minor axes defined through the spatial distribution of bound satellite galaxies in the simulation. The orange squares represent the best-fitting versus true mass within the half-mass radius of tracer satellites, $M(<r_\mathrm{half})$. The red dashed diagonal line marks ``$y=x$'' to guide the eye. Overall, \textsc{jam} gives a reasonable prediction of both $M_{200}$ and $M(<r_\mathrm{half})$. The errorbars are calculated from the boundaries defined by those models whose log likelihood ratios are within 1-$\sigma$ to the log likelihood value of the best model, assuming $\chi^2$ distribution for the twice log likelihood variable.}
\label{fig:fit_vs_real}
\end{figure*}

\begin{figure} 
\includegraphics[width=0.49\textwidth]{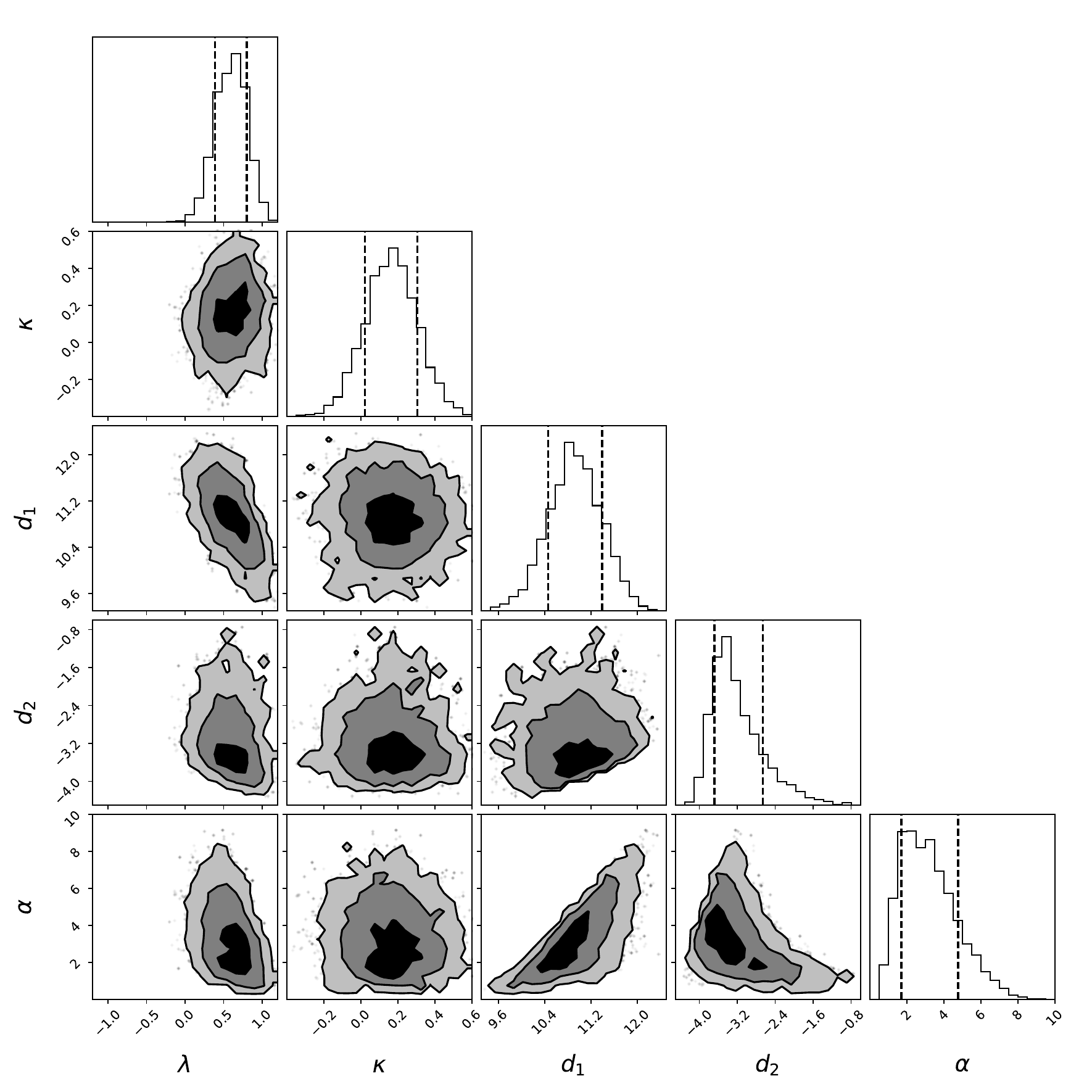}%
\caption{Error contours for one randomly selected galaxy cluster from TNG300. The black, dark gray, and gray contours are 1, 2, and 3-$\sigma$ confidence levels, respectively. The histograms on the right of each row show the 1-dimensional marginalized posterior distributions. The $x$ and $y$-axis ranges displayed here are chosen to be the same as Figure~\ref{fig:contour incl free} below. 
}
\label{fig:contour}
\end{figure}

Figure~\ref{fig:fit_vs_real} shows the \textsc{jam} predicted masses versus the truth in the simulation, for all 28 galaxy clusters selected from TNG300. Here the results are based on the case when the inclination angle, $incl$, is fixed. We will discuss the case when $incl$ is a free parameter later in Section~\ref{sec:incl_free}. We show the comparisons between the best-fitting and the truth for the total mass enclosed within the half-mass radius of tracer satellites, $r_\mathrm{half}$, and for the virial mass, $M_{200}$. Here $r_\mathrm{half}$ is defined as the projected radius, within which it contains half of the total bound tracer satellites projected within 2~Mpc. We denote the masses enclosed within $r_\mathrm{half}$ as $M(<r_\mathrm{half})$. Note the enclosed masses are defined in 3-dimension rather than in projection, according to the best-fitting potential model and the actual particle distributions in the original simulations. The best-fitting virial mass is calculated according to the best-fitting model density profile by \textsc{jam}. We first calculate the corresponding $R_{200}$ according to the best-fitting profile, and then calculate the integrated mass within $R_{200}$.

In general, the orange squares and the blue dots roughly distribute symmetrically around the red diagonal line, indicating reasonable and approximately ensemble unbiased mass constraints. There are small biases of -0.02 and 0.01 for $M(<r_\mathrm{half})$ and $M_{200}$. The mean scatters are $\sim$0.09~dex for $M(<r_\mathrm{half})$ and 0.15~dex for $M_{200}$. The readers can refer to the top row of Table~\ref{tbl:massbias} for a summary of these values.

According to the amounts of scatter, we can see the constraint on $M(<r_\mathrm{half})$ is better than that of $M_{200}$. This is consistent with the argument in many previous studies, which find that the mass within the half-mass radius of tracers is a sweet point, which can be constrained better than the masses within other radii \citep[e.g.][]{Wolf2010,2011ApJ...742...20W,2015MNRAS.453..377W,2017MNRAS.472.4786G,2020SCPMA..6309801W}. 
This is mainly due to the degeneracy between the two halo parameters ($d_1$ and $d_2$, $\rho_s$ and $r_s$, or $M_{200}$ and the concentration $c_{200}$). Perpendicular to the degeneracy direction, the constraint is the tightest, which corresponds to the amplitude of the potential at approximately the median radius of the tracer population. On the other hand, the constraint is the weakest along the degeneracy direction, which corresponds to the shape of the potential \citep[e.g.][]{2016MNRAS.456.1003H,2021MNRAS.505.3907L,Li2022}.

Figure~\ref{fig:contour} shows the error contours of different combinations of five model parameters ($\lambda$, $\kappa$, $d_1$, $d_2$, $\alpha$) for one randomly selected galaxy cluster. The black, dark gray, and gray regions show the 1, 2, and 3-$\sigma$ confidence intervals. As we have mentioned above, $d_1$ and $d_2$ are defined from the two halo parameters as $d_1\equiv \mathrm{log_{10}}(\rho_s^2r_s^3)$ and $d_2\equiv\mathrm{log_{10}}(\rho_s)$, and thus they are dimensionless and cover a much smaller range in log space than the original $\rho_s$ and $r_s$. The units we adopt for $\rho_s$ and $r_s$ are $\msun/\mathrm{pc}^3$ and pc in our calculations, and $d_2$ can be negative. The degeneracies between $d_1$, $d_2$ and $\alpha$ are prominent. The degeneracies between the rotation parameter, $\kappa$, and halo parameters ($d_1$, $d_2$ and $\alpha$) are very weak or absent, while there also exist some correlations between $\lambda$ and the three halo parameters, though not as strong as those among the halo parameters. Note for this galaxy cluster, $\kappa$ is positive but still close to zero, indicating weak rotations. For most of our galaxy clusters, the values of $\kappa$ are close to but not exactly zero, indicating galaxy clusters can have weak rotations, but they are not rotationally dominated systems. In the next subsection, we move on to investigate a few example density profiles and perform more detailed discussions.

\subsection{Example density and velocity dispersion profiles}

\begin{figure*} 
\includegraphics[width=0.95\textwidth]{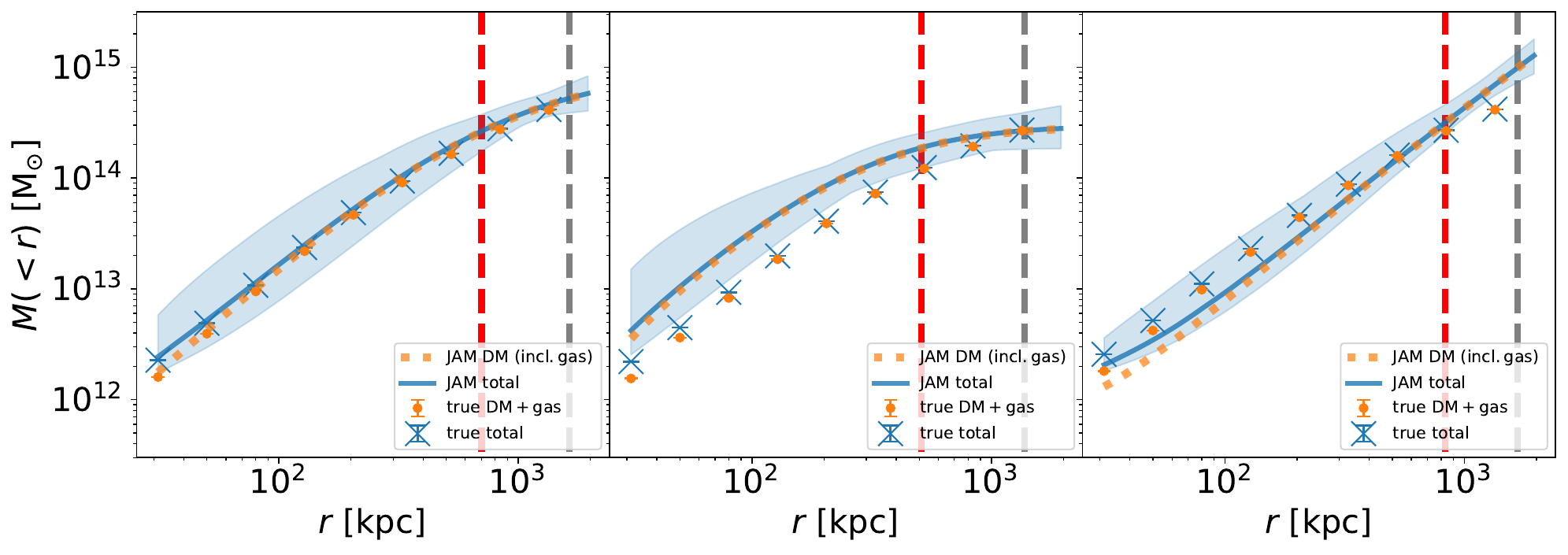}%
\caption{A comparison between the real and \textsc{jam} predicted density profiles for three example galaxy clusters from TNG300, and the three examples are shown in three different panels. The density profiles are calculated in 3-dimension, and $r$ in the $x$-axis label is the radius from the cluster center in 3-dimension. The true total density profile is represented by blue crosses with errorbars (comparable to the symbol size), and the blue solid curve is the \textsc{jam} prediction. The blue shaded region corresponds to the 1-$\sigma$ error by \textsc{jam}. The true dark matter $+$ gas density profiles are shown by orange dots, while the corresponding \textsc{jam} predictions are shown by the orange dotted curves. 
Note \textsc{jam} does not model the gas component separately, which is largely modeled within the dark matter component, so for fair comparisons, the orange dots correspond to the actual dark matter $+$ gas density profiles in the simulation. The true total mass density profile corresponds to everything in the simulation, including stellar, dark matter and gas, whereas the \textsc{jam} predicted total profile is the summation of the best constrained dark matter and stellar components, with the stellar component directly inferred by deprojecting the stellar surface density distribution. Errorbars of the true density profiles are calculated from the 1-$\sigma$ scatters of 100 bootstrap subsamples of particles in the galaxy clusters. The red and gray dashed vertical line mark the half-mass radii of the tracer satellites and $R_{200}$.}
\label{fig:profile}
\end{figure*}

\begin{figure*} 
\includegraphics[width=1.0\textwidth]{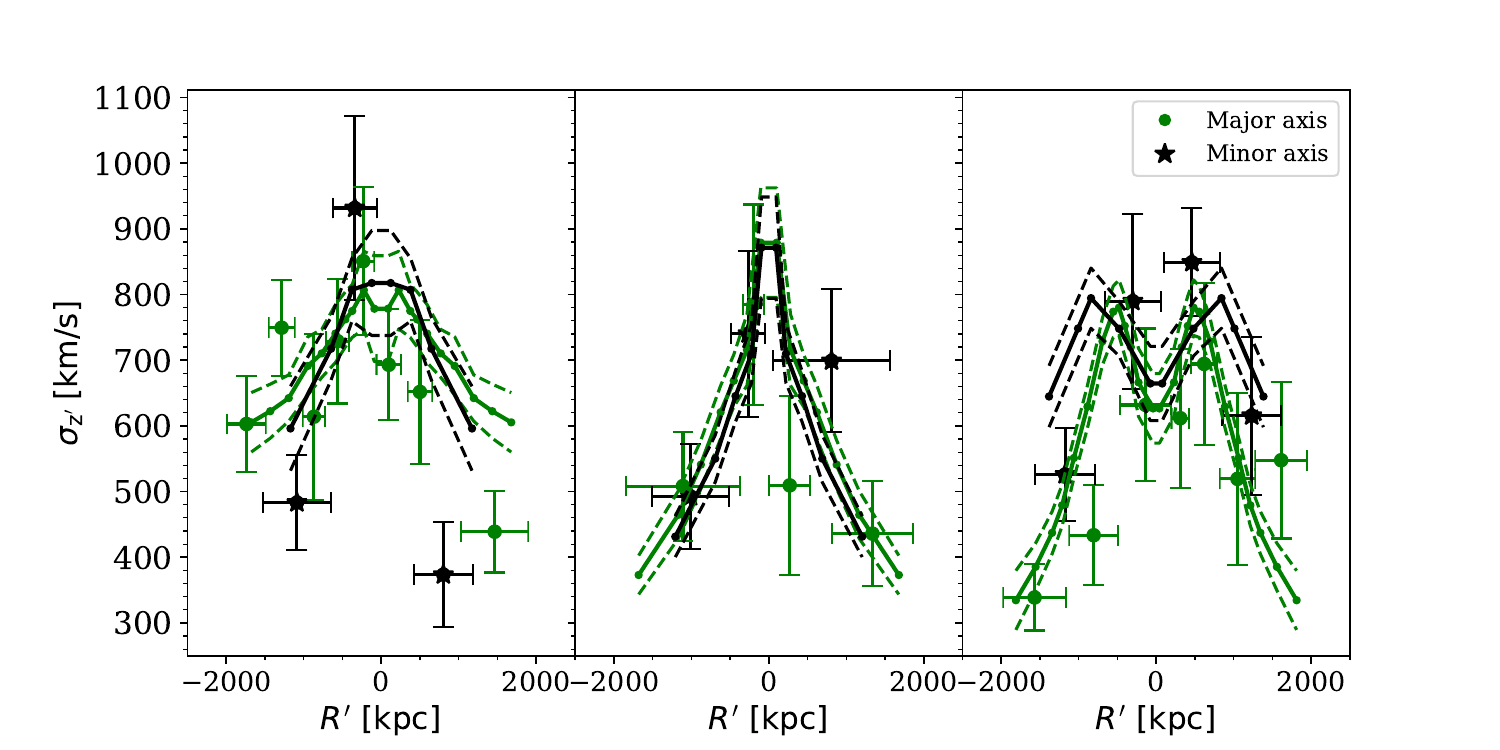}%
\caption{The line-of-sight ($z'$ component) velocity dispersion profiles of member satellites in three galaxy clusters, binned within sectors of $\pm45$ degrees to the major (green) and minor (black) axes of the cluster systems. Each bin contains 15 satellites. $R'$ indicates the projected distances to the cluster center in the corresponding sectors along the major or minor axes in the image plane. The $x$ and $y$ errors indicate the bin width and the 1-$\sigma$ scatters, respectively. Green and black solid curves show the best model predictions along the major and minor axes. The three galaxy clusters correspond exactly to the three clusters shown in Figure~\ref{fig:profile} above. }
\label{fig:velocity}
\end{figure*}

In this subsection, we further compare the best-fitting and true density and velocity dispersion profiles. Figure~\ref{fig:profile} shows the true and best-fitting matter density profiles for the dark matter $+$ gas (orange) components and for the total matter distribution (blue). This is shown for three representative galaxy clusters. At most of the radii, the blue crosses and solid curves are very close to the orange dots and dotted curves, except for the very inner regions, which is due to the contribution by the stellar component. Note again in our \textsc{jam} modeling, the gas component is not modeled separately. Instead, the hot gas is largely included in the dark matter component of the model, because they have similar distributions as the underlying dark matter in the simulation. In our analysis, the stellar component is directly deprojected from the stellar surface density distribution and is subdominant, and hence are not shown in Figure~\ref{fig:profile}.

In the left panel of Figure~\ref{fig:profile}, the difference between the blue crosses and the blue solid line is significantly smaller than the shaded errors, indicating the total density profile is very well recovered over the whole radial range. As we have explicitly checked, 8 out of the 28 systems ($\sim$29\%) in our galaxy sample belong to this case, i.e., the density profiles are very well constrained at almost all radii within $R_{200}$.  

In the middle panel, the best-fitting total profile more prominently deviates from the true total density profile at most of the radial range, and the total mass within the half mass radius of tracer satellites, $M(<r_\mathrm{half})$, is less well recovered. Though given the large statistical errors, the best constrained model and the truth still marginally agree with each other. About 9 galaxy clusters ($\sim$32\%) show similar trends as this middle panel, among which 5 show over estimates over most of the radial range, and 4 show under estimates over most of the radial range. 

In the right panel of Figure~\ref{fig:profile}, the best constrained total mass density profile is under estimated in the inner region, while over estimated in the outskirts. The best constrained and true profiles cross at approximately the half-mass radius of tracer satellites, as marked by the vertical red dashed line, leading to a good constraint on $M(<r_\mathrm{half})$. There are about 11 ($\sim$39\%) galaxy clusters in our analysis having their best-fitting inner and outer densities biasing from the truth in different directions, while maintaining a good recovery of $M(<r_\mathrm{half})$. Among them, 7 have under-estimated inner densities and over-estimated outer densities, while 4 have over-estimated inner densities and under-estimated outer densities. 

In previous studies, we have applied \textsc{jam} to dwarf galaxies in numerical simulations to recover their dark matter distributions \citep{2022ApJ...941..108W,2023ApJ...956...91W}. We find that contraction or infalling motions can cause deviations from steady states. Such infalling motions reduce the velocity dispersions in inner regions, resulting in under-estimated inner densities, and to maintain a good constraint on $M(<r_\mathrm{half})$, the outer densities are over estimated. On the other hand, global expansion motions such as gas outflows can cause over-estimated inner densities and under-estimated outer densities. 

Moreover, we have discussed in another study \citep{Li2022} that galaxy cluster systems having over-estimated $M(<r_\mathrm{half})$ often have large virial ratios\footnote{The virial ratio is defined as twice kinematical energy versus the potential energy of the system, calculated using all bound particles.}. Clusters with the highest virial ratio values are usually unrelaxed systems with high kinetic energy, which may be caused by major mergers or active mass accretion, and thus the kinetic energy is increased within a short time. \citep{Li2022} showed a tight correlation between the bias in $M(<r_\mathrm{half})$ and the system virial ratios.

In our current study, we have also investigated whether galaxy cluster systems with under/over-estimated inner densities and over/under-estimated outer densities have infalling/expansion motions in the tracer satellite population and in the gas and dark matter components. Unfortunately, we fail to see prominent correlations as in \citep{2022ApJ...941..108W}. Moreover, we have looked at the correlation between the bias in $M(<r_\mathrm{half})$ and the virial ratio, and fail to see prominent correlations. We think this is mainly limited by our small number of tracer satellite galaxies and the associated large statistical errors. Note in our current analysis, the number of tracer satellites ranges from 70 to slightly more than 350, whereas \citep{2022ApJ...941..108W} used at least 6,000 member star particles as tracers, and \citep{Li2022} in fact adopted dark matter particles in their simulations as tracers, instead of satellites or subhalos. The average number of tracers in galaxy clusters used by \citep{Li2022} is on the order of $10^5$.

We note that the contraction/infalling or expansion/outflow motions are not the only ways of causing the systems to deviate from steady states. The deviation from steady states can be caused by many other factors, such as the existence of massive and dynamically cold streams post major mergers, the perturbation by a massive companion satellite galaxy and the time evolution of the underlying gravitational potential. For the case shown in the middle panel of Figure~\ref{fig:profile}, we discovered a massive companion located behind the cluster along the line of sight, but passed our selection criterion along the line-of-sight direction. This is likely the reason causing the over estimate in $M(<r_\mathrm{half})$ for this system. For the other three systems having over-estimated $M(<r_\mathrm{half})$, we have identified one system having a massive companion projected just beyond 2~Mpc, but we fail to see similar existence of massive companions for the other two. However, we have tested our results by varying the magnitude gap when selecting our galaxy cluster systems. We find, a smaller magnitude gap in the selection, which means possible existence of more massive companions, would end up with more cases corresponding to the middle panel of Figure~\ref{fig:profile}.

Moreover, \textsc{jam} assumes axis-symmetry, whereas realistic galaxy cluster systems from TNG300 are not ideally axis-symmetric. The deviation from the axis-symmetric assumption is also responsible for the biases in the mass profiles. We find that different choices of line-of-sight direction with respect to the minor axis of our cluster systems can lead to different results for individual systems. For one system belonging to the classification in the middle panel of Figure~\ref{fig:profile}, its major axis is more aligned with the line of sight, which is likely the cause for over-estimated masses at most of the radii. 

So far we have demonstrated a few typical cases of how the best-fitting density profiles deviate from the truth. However, what we directly fit are the velocity moments, instead of the density profiles. We thus show in Figure~\ref{fig:velocity} the true (symbols) and best-fitting (lines) velocity dispersion profiles in projected radial bins along the major (green) and minor (black) axes of three galaxy clusters, which correspond exactly to the three systems we show in Figure~\ref{fig:profile}. 

In all three panels, the best-fitting models agree with the truth reasonably. However, the true velocity dispersion profiles are not perfectly axis-symmetric, with prominent differences between the left and right-hand sides. However, \textsc{jam} is an axis-symmetric model. As a result, some of the asymmetric features are not possible to be ideally fit. For example, if looking at the velocity dispersion profiles in the left panel, the model tends to be higher than the actual velocity dispersions on the positive side of the $x$-axis and at large radii, and the difference is greater than the errorbars, while the negative side is better fit. We do not expect the axis-symmetric \textsc{jam} model to have an ideally good fit for this case.

\subsection{Constraints on the inclination angle}
\label{sec:incl_free}

\begin{figure*}
\includegraphics[width=0.5\textwidth]{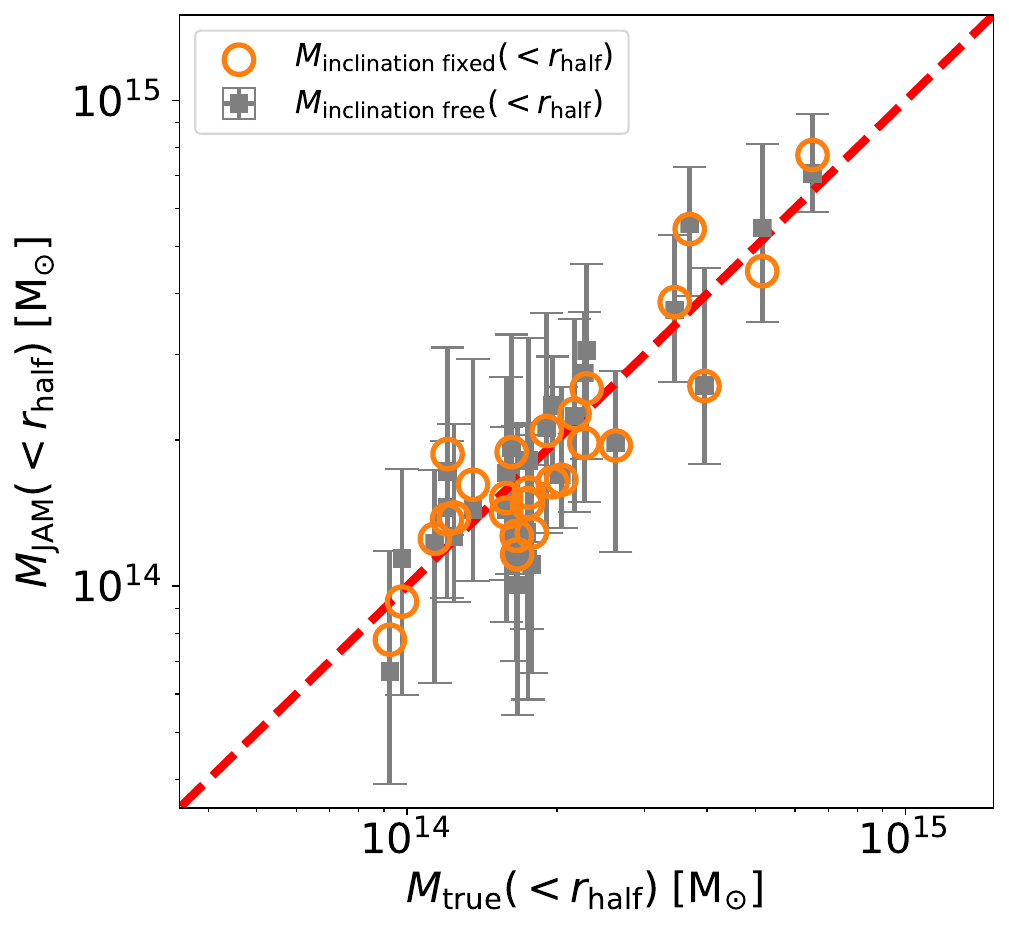}%
\includegraphics[width=0.5\textwidth]{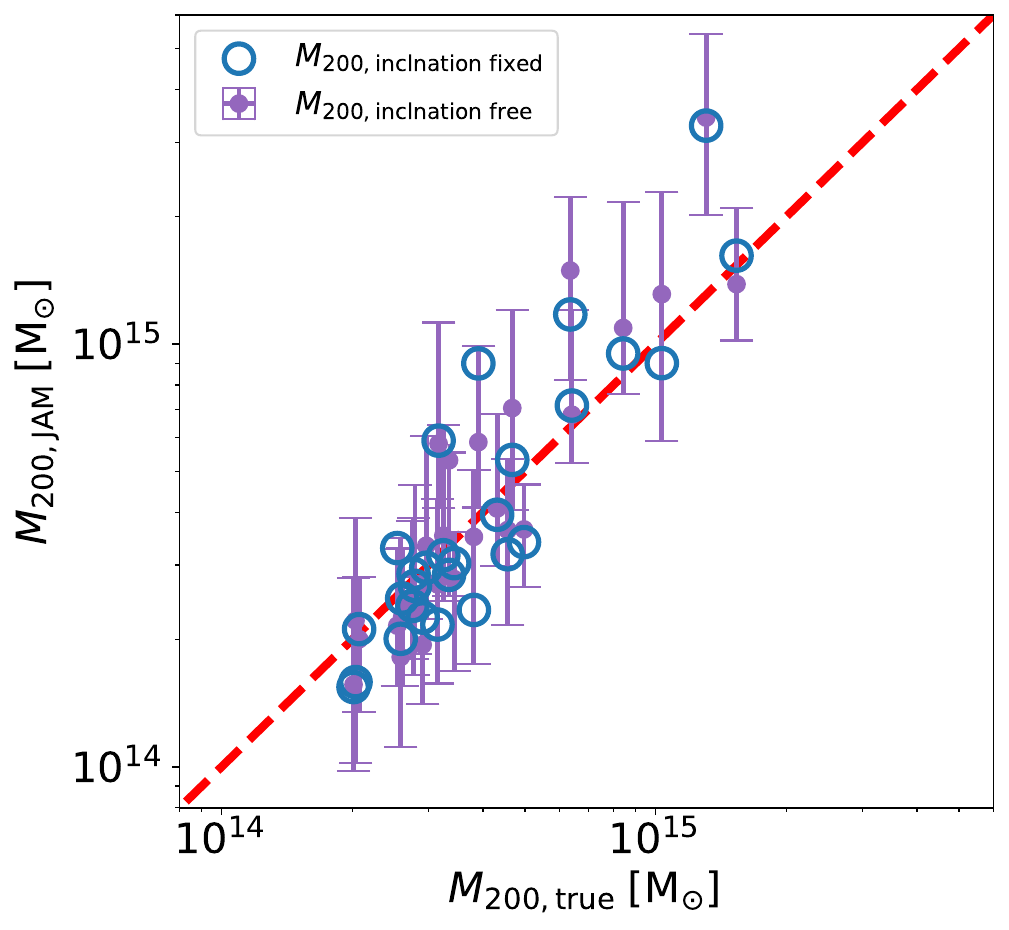}%
\caption{{\bf Left:} The best-fitting mass within the half-mass radius of tracer satellites ($M(<r_\mathrm{half})$) versus the truth for 28 galaxy clusters from TNG300. Here we set the inclination, $incl$, as a free model parameter, and the results are shown as gray squares with errorbars. Empty orange circles are repeats of the measurements in the previous Figure~\ref{fig:fit_vs_real}, when $incl$ is fixed to the truth in the simulation. Errorbars for the red circles are comparable to those for the orange squares, and are hence not repeatedly shown. {\bf Right:} Similar to the left plot, but for the virial mass ($M_{200}$). The blue circles are exactly the same as those in Figure~\ref{fig:fit_vs_real}.}
\label{fig:fit_vs_real_incl}
\end{figure*}

\begin{figure} 
\includegraphics[width=0.49\textwidth]{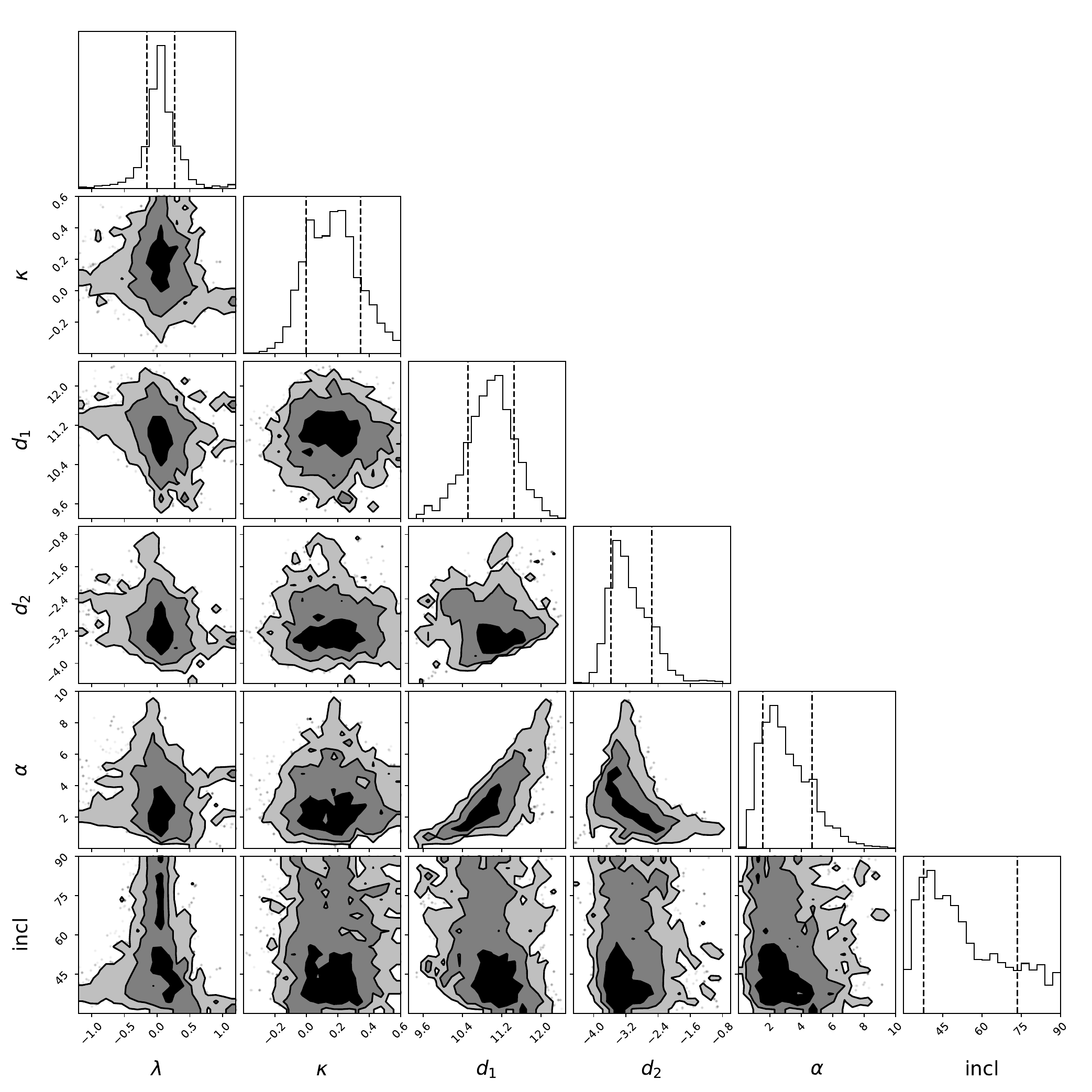}%
\caption{Similar to Figure~\ref{fig:contour}, but now the inclination angle, $incl$, is treated as a free model parameter.}
\label{fig:contour incl free}
\end{figure}

Our analysis in the previous subsections is achieved by fixing the inclination angle, $incl$, to the angle between the line of sight and the minor axis calculated from the spatial distribution of tracer satellites in TNG300. In real observation, we do not know $incl$ in advance. Thus from now on, we treat $incl$ as a free parameter in our modeling. Figure~\ref{fig:fit_vs_real_incl} shows a comparison between the cases when $incl$ is fixed or set free. For most of the cases, fixing $incl$ or setting it free leads to differences in best constrained $M(<r_\mathrm{half})$ and $M_{200}$ smaller than the errorbars. With $incl$ as a free parameter, the biases in $M(<r_\mathrm{half})$ and $M_{200}$ become -0.01 and 0.03~dex, and the scatters become 0.11 and 0.15~dex. The biases are not significantly different from the values when fixing $incl$ to the truth (-0.02 and 0.01~dex, see Section~\ref{sec:overall} above). The scatter in $M(<r_\mathrm{half})$ gets slightly larger, as compared to the value of 0.09~dex in $M(<r_\mathrm{half})$ of Section~\ref{sec:overall}. The readers can also refer to the top and middle rows of Table~\ref{tbl:massbias} for a summary of these values.

Figure~\ref{fig:contour incl free} shows the error contours for the same system as Figure~\ref{fig:contour}. The constraint on $incl$ is not tight, with a 1-$\sigma$ uncertainty of about 30~deg. Most of the other parameters do not show strong correlations with $incl$, except for $\lambda$, which shows some positive correlations with $incl$. As a result, the best constrained values of $\lambda$ differ more significantly between Figure~\ref{fig:contour} and Figure~\ref{fig:contour incl free}. The best constraints on the other model parameters agree well within 1-$\sigma$ either fixing $incl$ or setting $incl$ free.

The frequencies corresponding to the cases in the three panels of Figure~\ref{fig:profile} remain largely similar when $incl$ is a free parameter. So we do not repeatedly show the examples.

\subsection{The effect of foreground and background contamination}
\label{sec:contamination}

\begin{figure*}
\includegraphics[width=0.5\textwidth]{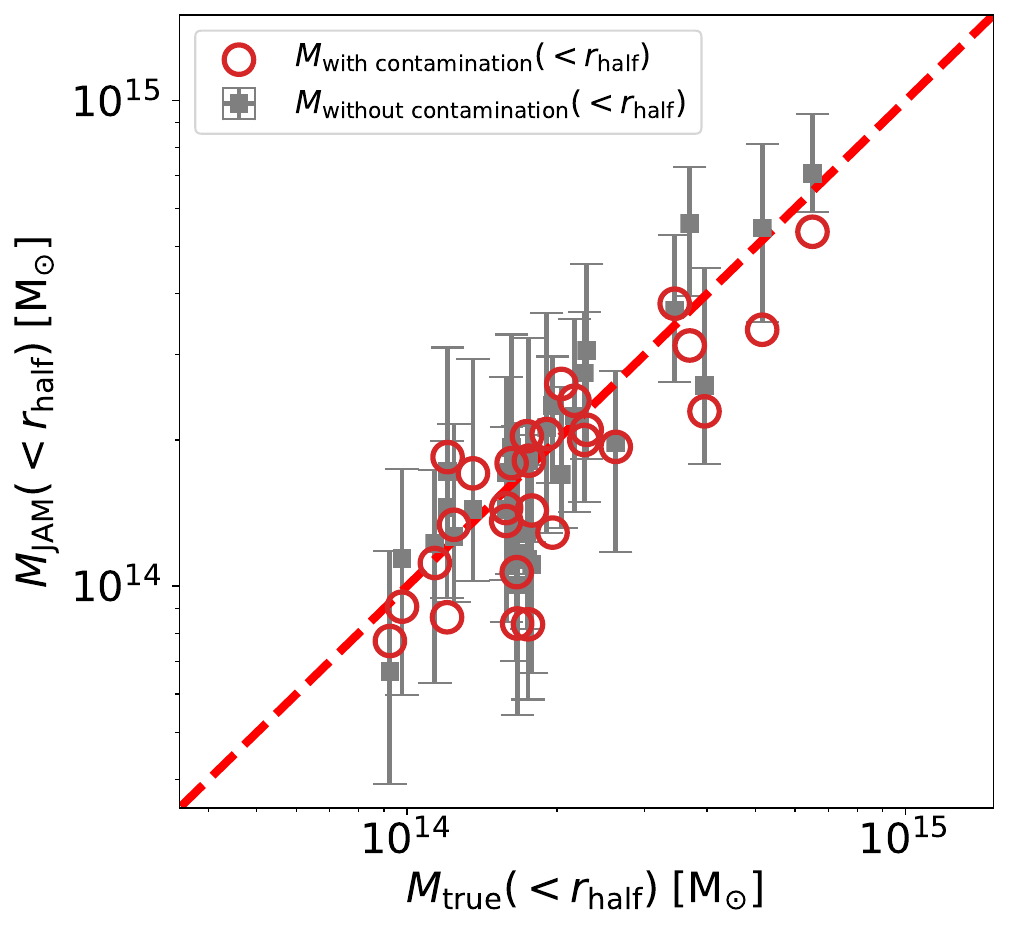}%
\includegraphics[width=0.5\textwidth]{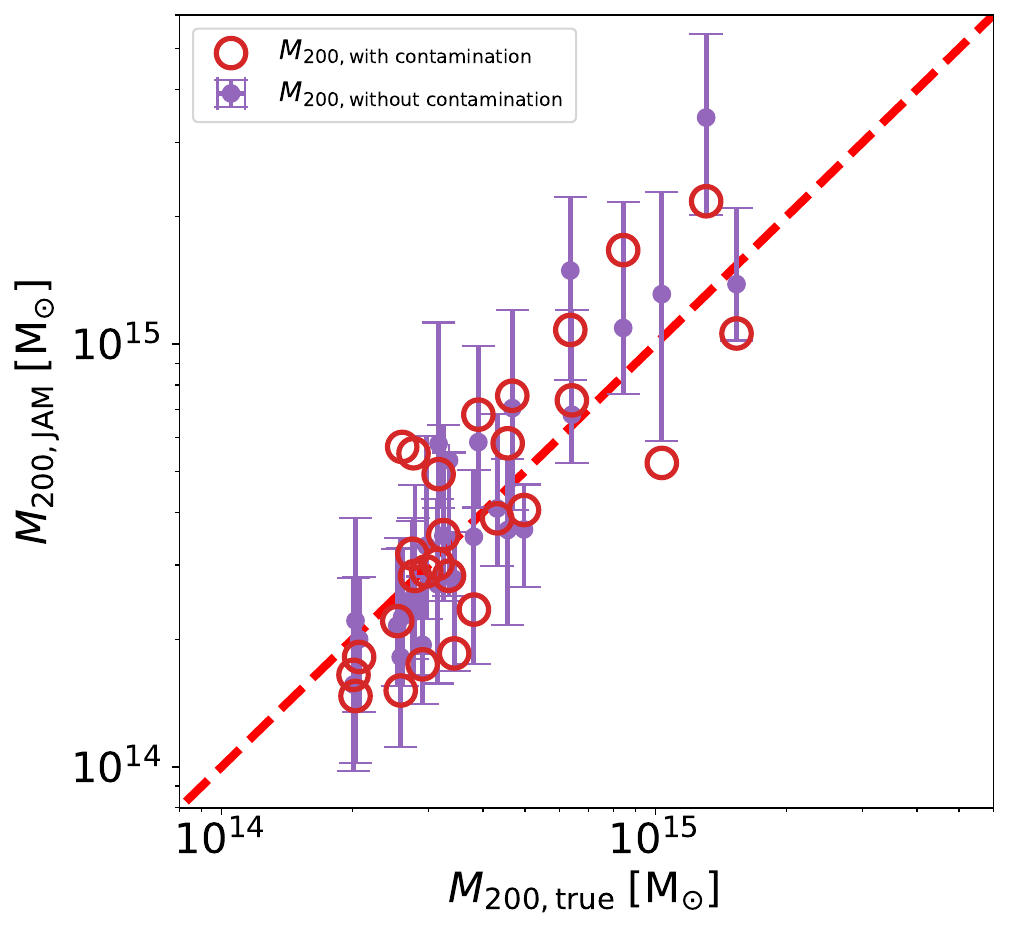}%
\caption{{\bf Left:} The best-fitting mass within the half-mass radius of tracer satellites ($M(<r_\mathrm{half})$) versus the truth for 28 galaxy clusters from TNG300. Gray squares are exactly the same as those in Figure~\ref{fig:contour incl free}. After selecting tracer satellites according to the line-of-sight velocity differences, we include $\sim$10\% contamination, and the corresponding results are shown as red empty circles. Errorbars for the red circles are comparable to those for the orange squares, and are hence not repeatedly shown. {\bf Right:} Similar to the left plot, but for the virial mass, $M_{200}$. The purple dots are exactly the same as those in Figure~\ref{fig:fit_vs_real}. The inclination angle, $incl$, is set as a free model parameter for results in both panels. We do not see prominent differences between the results before and after including contaminations.}
\label{fig:fit_vs_real_contami}
\end{figure*}

The results presented so far are based on using bound satellites as tracers. However, in real observation, satellite galaxies are selected in redshift space, which may suffer from contamination by foreground and background galaxies (purity), and true satellite galaxies might be missing (completeness).

In order to better mimic the selection of satellite galaxies in real observations, we project the TNG300 simulation box along the $Z$-axis, i.e., the line of sight, also defined as $z'$-axis in the observing frame. Tracer satellites are selected as those companions which are projected within 2~Mpc and with line-of-sight velocity differences with respect to the central galaxy smaller than 2,000~km/s. By selecting companion galaxies in this volume, we find on average 88\% true satellite galaxies in the simulation can be included, while there is only $\sim$11\% contamination by unbound galaxies. Note for results in this subsection, we treat $incl$ as a free parameter.

Figure~\ref{fig:fit_vs_real_contami} shows a comparison in the best recovered masses, based on true bound satellite galaxies and based on companions selected in projection as described above. Perhaps because of the reason that the fraction of contamination is as low as $\sim$11\%, the results before and after considering the contamination do not show significant differences for most systems. The associated biases and scatters for $M(<r_\mathrm{half})$ and $M_{200}$ are shown in the bottom row of Table~\ref{tbl:massbias}. Compared with the middle row when only bound satellites are used (no contamination), the biases and scatters only become slightly larger.

Our results in Figure~\ref{fig:fit_vs_real_contami} suggest that with the selection of companions projected in 2~Mpc and within 2,000~km/s along the line of sight, the contamination of fore/background can be controlled to be as low as 11\%, and the dynamical modeling outcome has slightly larger biases and scatters than using true bound satellites as tracers.

%%%%%%%%%%%%%%%%%%%%%%%%%%%%%%%%%%%%%%%%%%%%%%%%%%%%%%%%%
\begin{table}[!htbp]
\caption{A summary of biases and scatters in $M({<r_\mathrm{half}})$ and $M_{200}$. Throughout this paper, we first investigated the case when only bound satellites are used as tracers, with fixed $incl$ (Section~\ref{sec:overall}). We then moved on with the case of treating $incl$ as a free model parameter, but still using bound satellites as tracers (Section~\ref{sec:incl_free}). In the end, we select satellite galaxies in redshift space as tracers, with free $incl$ (Section~\ref{sec:contamination}). The three cases refer to the three rows of the table (see the text in the left column).}
\begin{center}
\begin{tabular}{c cc|cc}\hline
\hline
&\multicolumn{2}{c|}{Mass bias}&\multicolumn{2}{c}{Mass scatter}\\
 & \multicolumn{1}{c}{$M({<r_\mathrm{half}})$} & \multicolumn{1}{c|}{$M_{200}$}& \multicolumn{1}{c}{$M({<r_\mathrm{half}})$}&\multicolumn{1}{c}{$M_{200}$}\\ \hline
bound satellite &  &  &  &  \\ 
$incl$ fixed & { -0.02} &{ 0.01} & { 0.09} &{ 0.15}\\
\hline
bound satellite &  &  &  &  \\ 
$incl$ free & { -0.01} &{ 0.03} & { 0.11} &{ 0.15}\\
\hline
with contamination &  &  &  &  \\ 
$incl$ free & { -0.06} &{ 0.01} & { 0.12} &{ 0.18}\\
\hline
\hline
\label{tbl:massbias}
\end{tabular}
\end{center}
\end{table}
    
%%%%%%%%%%%%%%%%%%%%%%%%%%%%%%%%%%%%%%%%%%%%%%%%%%%%%%%%

\subsection{Fiber incompleteness and the dependence on flux limit}
\label{sec:fiber incompleteness}
\begin{figure*}
\includegraphics[width=0.9\textwidth]{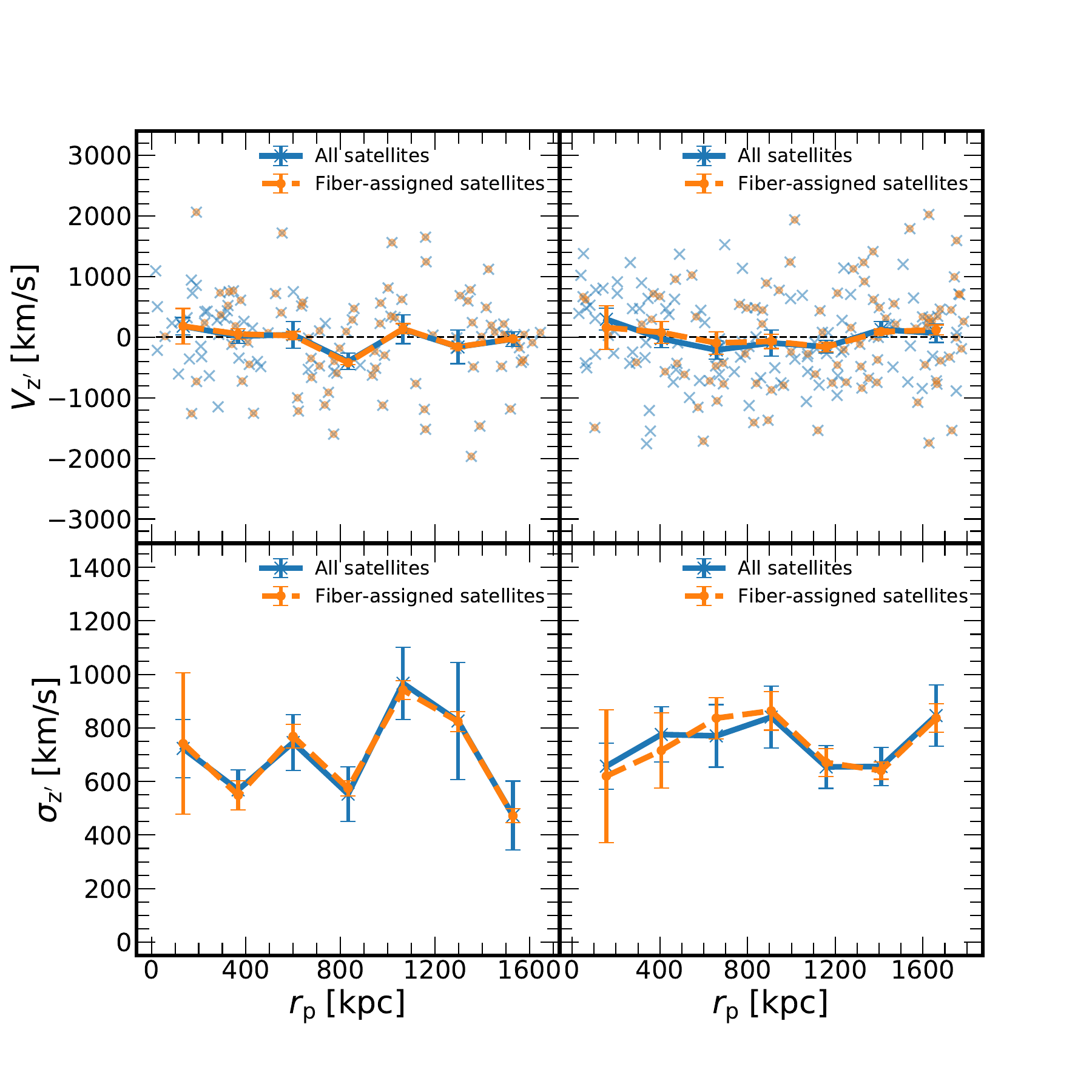}%
\caption{{\bf Top:} Upper panels show the line-of-sight velocity profiles based on satellites of two randomly selected galaxy clusters from the mock DESI BGS catalog, denoted as Cluster1 and Cluster2. The mean velocities are binned and calculated in projected circular annuli, and we use $r_p$ to denote the projected distance to the cluster center. The blue solid curves represent the mean velocity profiles based on the full set of bound companion satellites, and the orange dashed curves show the mean velocity profiles based on a subset of satellites with successful spectroscopic redshift measurements after incorporating fiber assignment and the redshift success rate in the mock. The blue crosses and orange dots correspond to individual satellites before and after accounting for fiber assignment. The profiles are derived from the mean of 32 times different fiber assignment tests and the errorbars represent the 1-$\sigma$ scatters. The errors of the blue solid curves are based on the 1-$\sigma$ scatters of 100 bootstrap subsamples, while they are calculated as the 1-$\sigma$ scatters of the 32 different fiber assignment tests for the orange dashed curves. {\bf Bottom:} Curves in the lower panels are similar, but show the line-of-sight velocity dispersion profiles for Cluster1 and Cluster2. All panels are based on the flux limit of $r<19.5$.}
\label{fig:fiber}
\end{figure*}

\begin{figure*}
\includegraphics[width=0.9\textwidth]{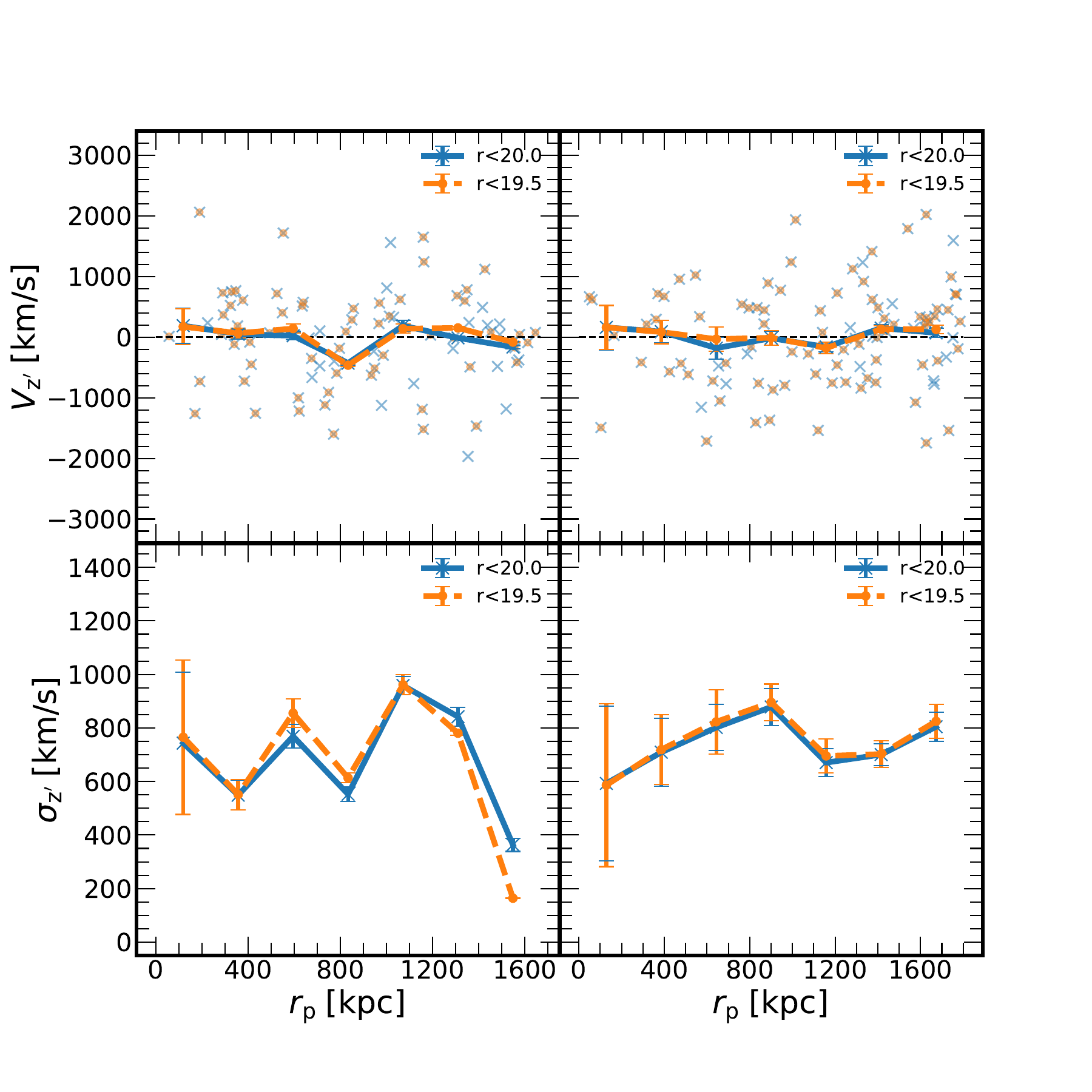}%
\caption{Similar to Figure~\ref{fig:fiber}, but now we compare the velocity (upper panels) and velocity dispersion (lower panels) profiles for two different flux limits ($r$-band flux limit of $r<20.0$ and $r<19.5$, in blue and orange as shown by the legend). The mean velocities and velocity dispersions are binned and calculated in projected circular annuli, and we use $r_p$ to denote the projected distance to the cluster center All curves in all panels are based on the fiber-assigned satellite galaxies. The profiles are derived from the mean of 32 times different fiber assignment tests and the errorbars represent the 1-$\sigma$ scatters of these tests.}
\label{fig:fiber2}
\end{figure*}

In addition to the projection effects and the contamination by foreground or background galaxies, real spectroscopic observations also suffer from fiber incompleteness and failures in redshift measurements, hence not all photometrically observed satellites would have spectroscopically measured line-of-sight velocities. The so-called fiber collision effect, that two fibers cannot be placed closer than a certain angular separation, is more severe in dense cluster regions. As having been evaluated by \cite{2019MNRAS.484.1285S}, with the DESI Bright Galaxy Survey fiber assignment strategy, the completeness fraction may be as low as 20\% for the worst cases in dense cluster regions. It is thus important to consider the effect of fiber incompleteness. 

In \textsc{jam}, the possible effects of fiber incompleteness come from two aspects: 1) the projected number density profile of tracer satellites is modified; 2) fiber incompleteness may change the velocity and velocity dispersion profiles. For 1), if one knows about the exact completeness fraction as a function of the projected distance to the cluster center (selection effect), corrections can in principle be made. For 2), the essential point is whether the subsample of satellites which have successful spectroscopic redshift measurements would alter the velocity and velocity dispersion profiles, compared with the full set of companion satellites. 

In order to check point 2) above, we look into a mock DESI bright galaxy survey (BGS) catalog based on the MXXL simulation \citep{2017MNRAS.470.4646S}. The readers can check Section~\ref{sec:mxxl} for details. Using the mock catalog, we investigate whether the velocity dispersion profiles are changed before and after considering the effect of fiber incompleteness. 

Figure~\ref{fig:fiber} shows the line-of-sight velocities and velocity dispersions as a function of the projected distance, $r_p$, to the cluster center, for two example galaxy clusters. We compare the velocity and velocity dispersion profiles before and after considering the effect of fiber incompleteness. The profiles are derived from the mean of 32 different random realizations of the fiber assignment algorithm, and the errorbars represent the 1-$\sigma$ scatters. The differences between the blue and orange curves are very small at all radii. Compared with the errors, we do not see any significant differences between the velocity dispersion profiles before and after incorporating the fiber incompleteness, indicating fiber incompleteness in DESI is unlikely to modify the velocity dispersions of observed satellite galaxies compared with the full sample of satellites, hence the dynamical modeling results are not likely to be affected either.

In addition to the effect of fiber incompleteness, we also examine the impact of flux limit on our analysis. The DESI BGS mock catalog includes 84 galaxy cluster systems within the DESI Year 5 footprint with a halo mass range of $14.3<\log_{10}M_\mathrm{halo}/\msun<15$ at $z<0.2$, that can have more than 100 satellites brighter than $r=19.5$ assigned with fibers. $r<19.5$ is the flux limit for the DESI BGS bright sample. The flux limit for the less complete BGS faint sample is $r<20.175$. Since the flux limit of the current version of the DESI BGS mock we are using is $r<20$ and if the flux limit is adjusted to $r<20$, there would be 165 galaxy clusters that can have more than 100 satellites above the flux limit. We perform tests to assess whether the exclusion of faint objects would affect the velocity and velocity dispersion profiles.

In Figure~\ref{fig:fiber2}, we present the radial velocity and velocity dispersion profiles for two representative clusters, subject to different flux limit cuts of their satellites ($r<19.5$ and $r<20.0$). The upper panels show the velocity profiles, where the blue and orange curves correspond to the flux limits of $r<20.0$ and $r<19.5$, respectively. We can see the orange curves are within the errorbars of the blue curves. The blue crosses and orange dots behind the curves represent the velocities of individual satellites with $r<20.0$ and $r<19.5$, respectively. The $V_{z^{\prime}}=0$ line goes well through both symbols, based on which we do not see significant selection biases. The lower panels show the velocity dispersion profiles. It can be seen that the orange curves are closely aligned with the blue curves. The results are thus unlikely to be significantly affected by the different choices of flux limits in DESI.

Our results indicate that the velocity and velocity dispersion profiles are not sensitive to the choice of flux limit in the DESI survey. Note as we have mentioned in Section~\ref{sec:mxxl}, the mock DESI BGS catalog we are using is by \cite{2017MNRAS.470.4646S}. The fiber assignment algorithm, the flux limit for the BGS faint sample and the number of passes are currently being updated for the latest version of the DESI BGS mock. These updates are not yet fully fixed, so we are focusing on the DESI BGS mock by \cite{2017MNRAS.470.4646S}. However, we expect these changes will not have a significant impact on the results presented in our analysis.

\section{Conclusions}
\label{sec:concl}

In this study, we investigate the performance of the Jeans Anisotropic Multi-Gaussian Expansion (\textsc{jam}) method, when it is applied to satellite galaxies in galaxy clusters to recover the underlying matter distribution. 28 galaxy cluster systems are selected from the cosmological and hydrodynamical TNG300-1 simulation (TNG300 in short), with halo mass of $\log_{10}M_\mathrm{200}/\msun>14.3$. 

The best constrained total matter density profiles by \textsc{jam} deviate from the truth in different ways. We divide the deviations into three different categories: 1) very good overall agreement with the true density profiles at all radii (29\%), with the deviations smaller than the errors of the true profiles; 2) over or under estimates at most radii (32\%) and 3) under/over estimates within the half-mass radius of tracer satellites ($r_\mathrm{half}$) and over/under estimates outside $r_\mathrm{half}$, which maintains a good prediction of $M(<r_\mathrm{half})$ (39\%). Most of the best constrained models are still consistent with the true profiles within 1-$\sigma$ statistical uncertainties of the model.

If only using true bound satellites as tracers and fixing the inclination parameter to the angle between the line of sight and the minor axis of satellite spatial distributions, the best constrained total mass within the half-mass radius of satellites, $M(<r_\mathrm{half})$, and the virial mass, $M_{200}$, have average biases of -0.02 and 0.01~dex, with average scatters of 0.09~dex and 0.15~dex. If treating the inclination as a free model parameter, the biases become -0.01 and 0.03~dex for $M(<r_\mathrm{half})$ and $M_{200}$, with mean scatters of 0.11 and 0.15~dex. The constraint on $M(<r_\mathrm{half})$ is tighter than that of $M_{200}$. 

If selecting tracer companions in redshift space, by requiring that their line-of-sight velocity differences are within $\pm$2,000~km/s to the cluster central galaxy, we can maintain a high completeness of 88\%, with the fraction of contamination by foreground and background galaxies as 11\%. The average biases are then -0.06~dex for $M(<r_\mathrm{half})$ and 0.01~dex for $M_{200}$, with mean scatters of 0.12 and 0.18~dex. 

We look into a mock DESI Bright Galaxy Survey (BGS) light-cone catalog, and find within the DESI Year 5 footprint, 84 galaxy cluster systems at redshift $z<0.2$ can have more than 100 satellite galaxies brighter than $r=19.5$ and with fiber assignments. If the flux limit is changed to $r<20$, there are 165 galaxy clusters that satisfy the selection. Hence it is promising to constrain the mass of galaxy clusters using satellite dynamics with future DESI data. Based on the mock DESI catalog, we test the effect of fiber incompleteness and the dependence on the survey flux limit and see no significant changes in the velocity and velocity dispersion profiles within $R_{200}$. Thus selection effects brought in by fiber assignments and survey flux limits are unlikely to affect the dynamical modeling outcomes.

We conclude it is promising to apply \textsc{jam} to satellite galaxies in galaxy clusters in on-going and future deep spectroscopic surveys, to constrain the underlying density profiles and total mass of individual cluster systems. Our quoted amounts of biases and scatters can be used as corrections to future observational constraints.

\acknowledgments

This work is supported by NSFC (12273021, 12022307), the National Key R\&D Program of
China (2023YFA1605600, 2023YFA1605601, 2023YFA1607800, 2023YFA1607801), the China Manned Space (CSST) Project with No. CMS-CSST-2021-A02 and No. CMS-CSST-2021-A03, the National Key Basic Research and Development Program of China (No.~2018YFA0404504), 111 project (No.~B20019) and Shanghai Natural Science Foundation (No. 19ZR1466800). We thank the sponsorship from Yangyang Development Fund. Z-L acknowledges the funding from the European Unions Horizon 2020 research and innovation programme under the Marie Skodowska-Curie grant 101109759 (“CuspCore”). L-Z acknowledges the support from the National Natural Science Foundation of China under grant No. Y945271001, and CAS Project for Young Scientists in Basic Research, Grant No. YSBR-062. R-S is grateful for useful discussions about light-cone mocks with Zhenlin Tan, and grateful for the discussions on Jean's equations with Cong Liu, Tianye Xia and Zhao Chen. R-S finalized calculations under very close supervisions by the corresponding author, Z-L and a few other coauthors. The paper writing of R-S's papers is almost all finished by the corresponding author. 

The computation of this work is carried out on the \textsc{Gravity} supercomputer at the Department of Astronomy, Shanghai Jiao Tong University.
This work has made extensive use of the \textsc{python} packages -- \textsc{ipython} with its \textsc{jupyter} 
notebook \citep{ipython}, \textsc{numpy} \citep{numpy} and \textsc{scipy} \citep{scipy}. All the figures 
in this paper are plotted using the python matplotlib package \citep{Matplotlib}. This research has made use of 
NASA's Astrophysics Data System and the arXiv preprint server. 

We gratefully thank the anonymous referee for his/her careful reading of the paper and useful comments, which have helped to significantly improve the paper.

\clearpage

\bibliography{paper}{}
\bibliographystyle{aasjournal}

\clearpage

%\appendix
%\section{}

%\begin{figure*} 
%\includegraphics[width=0.95\textwidth]{figures/image15gas.pdf}%
%\caption{The projected galaxy image corresponding to the galaxy cluster in the right panel of Figure~\ref{fig:profile}, %with the projection direction $X$, $Y$ and $Z$-axes of the TNG300 simulation box, respectively.}
%\label{fig:image}
%\end{figure*}

%\begin{figure} 
%\includegraphics[width=0.5\textwidth]{figures/velocity.pdf}%
%\caption{Radial velocities of baryon particles within the galaxy cluster in the right panel of Figure~\ref{fig:profile}. he mean velocity profile (black solid curve) and its 1-$\sigma$ uncertainty (black band) are computed
%from the baryon particles. The blue dashed line is the circular velocity calculated from $\sqrt{GM_{200}/R_{200}}$ and the black dashed line is the true $R_{200}$. }
%\label{fig:velocity}
%\end{figure}

\end{document}